\documentclass[pra,twocolumn,superscriptaddress,aps,longbibliography,preprintnumbers]{revtex4-2}
\usepackage{hyperref}
\hypersetup{
	colorlinks=true,
	citecolor=magenta,
	linkcolor=blue,
	urlcolor=blue}
\usepackage{graphicx}
\usepackage{amsmath}
\usepackage{physics}
\usepackage{amsfonts}
\usepackage{amssymb}
\usepackage{mathtools}
\usepackage{float} 
\usepackage{bm}
\usepackage{siunitx}
\usepackage{xfrac}
\usepackage[normalem]{ulem}
\usepackage{orcidlink}  
\usepackage{braket}% Dirac braket
\usepackage{cases}% Numbered cases

\usepackage{siunitx}

\usepackage[T1]{fontenc}

\begin{document}

%\title{Stochastic geometry and universality beyond the Kibble-Zurek mechanism in a newborn spinor Bose-Einstein condensate}
%\title{Universality beyond the Kibble-Zurek mechanism of point topological defects in a newborn coherently coupled Bose gas}

% \title{Universal statistics of topological defects in a newborn coherently coupled Bose gas}

% \title{Universality beyond the Kibble-Zurek mechanism in a newborn coherently coupled Bose-Einstein condensate}

% \title{Universality beyond the Kibble-Zurek mechanism in the Bose-Einstein condensation of coherently coupled Bose gases}
\title{Universality beyond the Kibble-Zurek mechanism in the condensation of coherently coupled Bose gases}

\author{Subhadeep Patra\orcidlink{0009-0009-2041-8539}}
\email{d23222@students.iitmandi.ac.in}
\affiliation{School of Physical Sciences, Indian Institute of Technology Mandi, Mandi-175005 (H.P.), India}
\author{Paolo Comaron\orcidlink{0000-0002-4492-4374}}
\email{paolo.comaron@cnr.it}
\affiliation{CNR NANOTEC, Institute of Nanotechnology, Via Monteroni, Lecce, Italy}
\author{Arko Roy\orcidlink{0000-0003-4459-2880}}
\email{arko@iitmandi.ac.in}
\affiliation{School of Physical Sciences, Indian Institute of Technology Mandi, Mandi-175005 (H.P.), India}

\begin{abstract}
We study the universal spatial statistics of point-like topological defects formed during the nonequilibrium condensation of a coherently coupled Bose gas using the stochastic projected Gross--Pitaevskii equation. The symmetry-breaking transition is driven by a linear quench of the chemical potential, leading to stochastic vortex nucleation in the individual condensate components. When the two components are considered together, these elementary defects may combine across components to emerge as composite topological defects known as full quantum vortices. Beyond the mean defect density predicted by the Kibble--Zurek mechanism (KZM), we investigate the spatial organization of both the elementary and composite defects and show that their positions are well described by a Poisson point process, revealing a universal stochastic geometry. This universality is further described through Voronoi tessellation, whose cell-area statistics follow Poisson--Voronoi predictions. We also introduce the spatial form factor for characterizing the vortex configurations and demonstrate the emergence of a characteristic dip--ramp--plateau structure. Our results establish universal stochastic geometry of topological defects beyond conventional Kibble--Zurek scaling and identify it as a fundamental feature of nonequilibrium condensation in coherently coupled Bose gases.
\end{abstract}
\maketitle

\section{Introduction}
Symmetry-breaking phase transitions and their rich dynamics have been extensively studied across a wide range of systems, including cosmology~\cite{Kibble_1976, kibble_1980, Morikawa_1995}, condensed matter physics~\cite{Buerle1996, Ruutu1996, Monaco_2009}, and ultracold atoms~\cite{Zurek1985, Ueda10, StamperKurn13, Campo_2014, Anquez_2016, Weiler2008, Navon_2015, Wu_2017}. A particularly striking aspect of these transitions is the spontaneous formation of topological defects~\cite{Dodd_1998, cruz_2007}, which provide valuable insights into the underlying symmetry-breaking process and a powerful route to identifying the universality class of the transition~\cite{Odor_2004, Chen_2019, Gavassino_2024}. 
The formation and subsequent dynamics of these topological defects after a continuous phase transition are well described by the Kibble-Zurek mechanism (KZM), originally proposed by Kibble in the context symmetry-breaking phase transition in the early universe~\cite{Kibble_1976}. 
% In his seminal work, Kibble predicts the possible emergence of cosmic strings and domain structures. 
Motivated by the possible formation of cosmic
strings and domain structures during cosmological evolution, Zurek extended these ideas to condensed matter systems undergoing continuous phase transitions ~\cite{Zurek1985,Zurek_1996}. Since then, the KZM has been studied in a wide range of systems for both classical and quantum phase transitions, such as cosmic microwave background~\cite{Bevis_2008}, superconducting films~\cite{Maniv_2003, Maegochi_2022}, liquid crystals~\cite{Chuang_1991}, colloidal monolayers~\cite{Deutschlander_2015}, Josephson junctions~\cite{Carmi_2000}, exciton-polaritons~\cite{zamora2020}, and quantum simulators~\cite{Keesling2019, Bando_2020}.

In a continuous phase transition, equilibrium scaling laws describe the onset of spontaneous symmetry breaking~\cite{HohenbergHalperin1977}.
For a control parameter $\varepsilon = (\lambda_{c} - \lambda)/\lambda_{c}$, which measures the proximity of the critical point $\lambda_{c}$, the correlation length $\xi$ and relaxation time $\tau$ obey the universal scaling laws given by,
\begin{equation}
    \xi(\varepsilon) = \frac{\xi_{0}}{|\varepsilon|^\nu}, \qquad
    \tau(\varepsilon) = \frac{\tau_{0}}{|\varepsilon|^{z\nu}},
    \label{tau_eq}
\end{equation}
where $\nu$ and $z$ are the static and dynamical critical exponents and $\xi_{0}$ and $\tau_{0}$ are system dependent constants. For a linear quench across the critical point, \(\varepsilon=t/\tau_Q\), performed from an initial time to a final time, the KZM predicts the existence of a characteristic freeze-out time \(\hat t\) at which the system can no longer adiabatically follow the changing control parameter~\cite{delCampo2013}.
%
% The KZM predicts that, for a linear quench across the critical point, there exists a characteristic time scale known as the freeze-out time $\hat{t}$. 
After which the system can no longer remain adiabatic and begins to respond to the changing control parameter $\varepsilon = t / \tau_Q$, where $t \in (t_i, t_f)$, which matches the relaxation time $\tau \sim |t| \sim \hat t$~\cite{delCampo2013}.
The resulting freeze-out time and associated correlation length scale universally with the quench time \(\tau_Q\) as
% Here, the freeze-out time $\hat t$ and the associated length-scale are predicted to have a universal scaling with the finite quench time $\tau_{Q}$ given by
\begin{equation}
    \hat t = (\tau_0 \tau_Q^{z \nu})^\frac{1}{1+z\nu}, \qquad 
    \hat \xi = \xi[\varepsilon(\hat t)] = \xi_0 \bigg(\frac{\tau_Q}{\tau_0} \bigg)^{\frac{\nu}{1+z\nu}}, 
    \label{xi}
\end{equation}
where the scaling exponents are determined solely by the equilibrium critical exponents~\cite{Zurek_2009, Campo2021}. The characteristic length scale \(\hat\xi\) sets the average separation between independently formed domains and consequently determines the density of the topological defects~\cite{HohenbergHalperin1977} generated during phase transition, 
\begin{equation}
    \rho_{\rm KZM}= \frac{1}{\hat \xi^{D-d}} = \frac{1}{\xi_0^{D-d}}\bigg(\frac{\tau_0}{\tau_Q}\bigg)^{\frac{(D-d)\nu}{1+z\nu}},
    \label{defect_KZ}
\end{equation}
where $D$ is the spatial dimension of the system and $d$ is the dimension of the topological defect. Thus, the defect
density exhibits a universal power-law dependence on the
quench time $\tau_Q$, independent of microscopic details. 
% The power law exponent $\nu/(1+z\nu)$ of Eq.~\eqref{xi} describing only by the equilibrium critical exponents. The value of the critical exponents decides the universality class of the system for which microscopic details are not important. 
% In scaling hypothesis, the universal dynamical behavior is characterizes by the correlation length~\cite{HohenbergHalperin1977}. 
For a spatially homogeneous system with $z = 2$ and $\nu=1/2$ in the mean-field regime~\cite{damski_2010, su_13}, Eq.~\eqref{defect_KZ} predicts $\rho_{\rm KZM} \propto \tau_Q^{-1/2}$ for point defects ($d = 0$) in two dimensions ($D = 2$). 
\begin{figure*}[!htbp]
    \centering
    \includegraphics[width=0.6\linewidth]{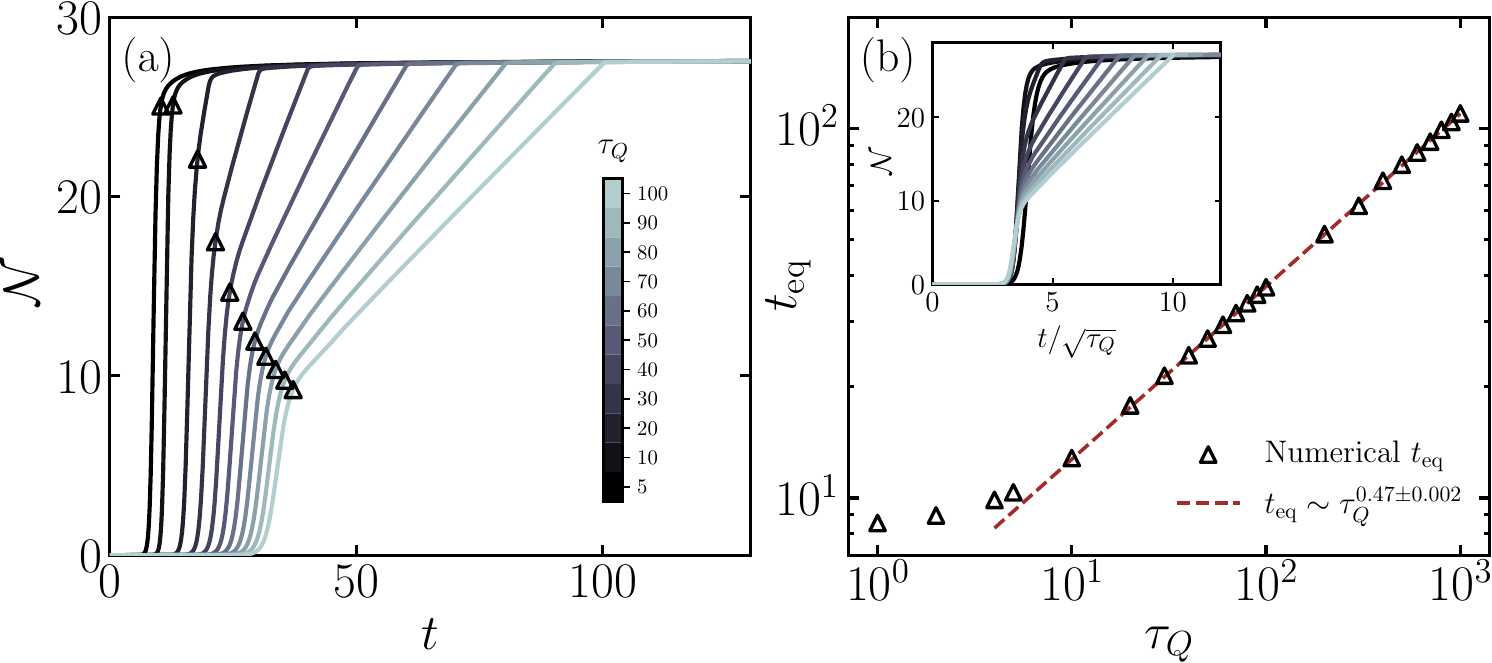}
    \caption{Characterization of the equilibration time $t_{\rm eq}$. 
(a) Scaled condensate norm $\mathcal{N}(t)$ as a function of time for different quench times $\tau_Q$ ranging from 5 to 1000. Each curve is averaged over $\mathcal{R}=100$ stochastic trajectories. Black triangles mark the equilibration time $t_{\rm eq}$. 
(b) Equilibration time $t_{\rm eq}$ as a function of the quench time $\tau_Q$. Fits in the range $\tau_Q=30$--1000 show good agreement with Kibble--Zurek scaling, $ t_{\rm eq} \sim \tau_Q^{0.47 \pm 0.002}$. 
% Error bars denote one standard deviation over $\mathcal{R}=100$ noise realizations. The error bars are very small; therefore, a contrasting color is used to enhance their visibility.
Inset: Growth of the condensate norm $\mathcal{N}$ versus the scaled time $t/\sqrt{\tau_Q}$. For $t < t_{\rm eq}$, the curves collapse onto a universal form for all $\tau_Q$ except the fastest quench $\tau_Q=5$, consistent with Kibble--Zurek behavior.
% \SP{this $t_{\rm eq}$ is the final time after shifting. We do not need to discuss the shifting procedure in the main text, since all the details are given in the Appendix.}
%\paolo{c seems to be not given}
}
    \label{t_eq}
\end{figure*}
Relevant to the present work, the formation and dynamics of vortices have been extensively investigated in a wide range of physical systems, including ultracold atomic gases~\cite{Weiler2008,Richaud_2020,Scherer_2007,Patra_2026}, superconductors~\cite{Hu_2024,chen_2024}, nanomagnets~\cite{Stebliy_2017}, exciton-polariton condensates~\cite{Sanvitto2010,panico2023}, and Josephson junctions~\cite{Xhani_2020}. Particular attention has been devoted to their universal nonequilibrium formation and evolution within the framework of the Kibble--Zurek mechanism (KZM)~\cite{Saito_2007,Ulm13,Pyka13,Donadello_2016,Reichhardt_2022,Maegochi_2022,Thudiyangal_2024,Liu2018}. Bose mixtures~\cite{Trippenbach_2000,Hofmann_2014} provide an especially rich setting for exploring such nonequilibrium dynamics owing to the interplay between multiple superfluid components~\cite{Matthews_1999,Patra_2026}. Unlike single-component condensates, multicomponent systems support a broader spectrum of topological excitations. In particular, coherent coupling between two condensate components can bind elementary vortices through a relative-phase domain wall, giving rise to composite topological defects known as vortex molecules~\cite{son_02,kasamatsu2004,su_13}. This makes coherently coupled Bose mixtures an attractive platform for investigating the emergence and universal dynamics of topological defects beyond the single-component paradigm. Beyond vortex dynamics, non-equilibrium behavior in mixtures of Bose-Einstein condensates (BECs) has also been investigated with particular emphasis on domain formation~\cite{De_2014}, their universality via KZM~\cite{Sabbatini_2012, Xu2016, Jiang_2024}, and subsequent coarsening dynamics~\cite{Singh_2023, Jiang_2024}.

Using the stochastic projected Gross-Pitaevskii equation (SPGPE) formalism (see Sec.~\ref{spgpe_frame}), we first investigate KZM universality of topological defect formation in Sec.~\ref{defect_kzm}.
Since the universality beyond mean defect density, particularly in coherently coupled Bose gases, remains largely unexplored, we go beyond conventional Kibble--Zurek scaling through spatial full counting statistics and investigate the universal spatial statistics of elementary and composite topological defects. Specifically, we show that the defect configurations are accurately described by a homogeneous Poisson point process (PPP), with a density determined by Kibble--Zurek scaling, and characterize the resulting stochastic geometry through Poisson--Voronoi tessellation (Sec.~\ref{Stochastic_geometry}). As a complementary probe of spatial correlations, we introduce the spatial form factor (SFF) in Sec.~\ref{sff_label}, drawing an analogy with the spectral form factor widely used in the study of integrable and quantum-chaotic systems. Together, these approaches reveal a universal stochastic geometry underlying defect formation that extends beyond the conventional Kibble--Zurek description based solely on defect density.

\section{Theoretical framework and quenching protocol}
\label{spgpe_frame}
To simulate symmetry-breaking dynamics in a newborn homogeneous Bose--Einstein condensate formed in two coherently coupled hyperfine states, $|1\rangle \equiv |\uparrow\rangle$ and $|2\rangle \equiv |\downarrow\rangle$, we employ the stochastic projected Gross--Pitaevskii equation (SPGPE) framework~\cite{Stoof_1999, Stoof2001, Gardiner_2003, Blakie2008, Proukakis2011, Rooney_2013, Ota_2018, roy_2021, Sivasankar_2026}, where the two components are linked by a Rabi coupling $\Omega$ that induces coherent population transfer between the states~\cite{Pethick_2008, recati_2021, farolfi_2021, Sunilkumar_2026, Sivasankar_2026}.
% To simulate the dynamics of symmetry breaking in a newborn homogeneous condensate with coherent (Rabi) coupling between two hyperfine states~\cite{Pethick_2008, recati_2021}, we employ the stochastic (projected) Gross--Pitaevskii equation (SPGPE) framework~\cite{Stoof_1999, Stoof2001, Gardiner_2003, Blakie2008, Proukakis2011, Rooney_2013, Ota_2018, roy_2021}. 
The coupled SPGPEs for a two-dimensional system are
\begin{equation}
\begin{split}
    i \hbar \frac{\partial}{\partial t} \psi_j (\mathbf{x},t) &= \hat{{\mathcal{P}}} \Bigg\{ (1 - i\gamma) \Bigg[ \Bigg( - \frac{\hbar^2 \nabla^2}{2 m} + g |\psi_j (\mathbf{x}, t)|^2 \\
    &+ g_{12} |\psi_{3 - j}|^2  - \mu(t) \Bigg) \psi_j + \Omega \psi_{3-j} \Bigg] + \eta_j (\mathbf{x},t) \Bigg\},
\end{split}
\label{sgpe}
\end{equation}
where $\mathbf{x}\equiv (x,y)$, $j=1,2$ labels the two spin components, and 
$\hat{\mathcal{P}}$ projects the dynamics onto the low-energy $\mathbf{C}$-field region 
below the cutoff $\epsilon_{\rm cut}$. The $\mathbf{C}$-fields are coupled to a thermal 
reservoir (the $\mathbf{I}$-region), which induces dissipation with strength $\gamma$ 
and complex Gaussian noise $\eta_j(\mathbf{x},t)$ satisfying the fluctuation--dissipation relation
\begin{equation}
    \langle \eta_i(\mathbf{x}, t) \eta_j^*(\mathbf{x}', t') \rangle 
    = 2\hbar\gamma k_B T \,\delta(\mathbf{x} - \mathbf{x}')\delta(t - t')\delta_{ij}.
\end{equation}
The cutoff is chosen as $\epsilon_{\rm cut}=k_B T\ln 2+\mu$, ensuring that the 
$\mathbf{C}$-region contains the macroscopically occupied low-energy modes, while 
higher-energy modes form the $\mathbf{I}$-region assumed to remain in local equilibrium 
at temperature $T$ and chemical potential $\mu$
~\cite{Blakie2008,Proukakis_2008,Comaron_2019,Larcher2018,Liu_2020,Rooney_2010}. This choice guarantees that the mean occupation of the modes below $\epsilon_{{\rm cut}}$ is larger than unity. The precise value of the cutoff is not crucial, as long as it belongs to a reasonable range. The same choice of the cutoff has been earlier used to validate experimental results for single component~\cite{ota_18,Comaron_2019,fabrizio_18} and two-component~\cite{roy_2021,roy_2023} condensates in similar configurations.
The parameter $\gamma$ controls the rate at which the $\mathbf{C}$-field exchanges particles 
and energy with the reservoir and hence sets the relaxation timescale towards equilibrium. While it has been shown that varying $\gamma$ would affect the long time universal single-component dynamics of topological defects in the ordered phase~\cite{groszek2021}, 
scaling behavior is expected to be insensitive to the precise value of $\gamma$ within the parameter ranges considered~\cite{Kasamatsu_2003, damski_2010}.
%
% Varying $\gamma$ only rescales relaxation times without affecting the critical dynamics. The universal features of the 
% nonequilibrium dynamics and scaling behavior are independent of the precise value of $\gamma$~\cite{Kasamatsu_2003, damski_2010}. 
% \paolo{well we found this is not actually true \cite{groszek2021} so this may need to be rephrased. maybe:}.

% \paolo{While it has been shown that varying $\gamma$ would affect the long time-dynamics of topological defects in the ordered phase~\cite{groszek2021}, 
% scaling behavior is expected to be insensitive to the precise value of $\gamma$ within the parameter ranges considered~\cite{Kasamatsu_2003, damski_2010}.}\SP{you are right. timescale of the scaling behavior in our case is too short for the dissipative effects associated with $\gamma$ to become significant.}
 % We study the formation of a newborn Bose--Einstein condensate in a coherently coupled two-component mixture, 
% with internal states $1 (|\uparrow\rangle)$ and $2 (|\downarrow\rangle)$ coupled by a Rabi term $\Omega$, 
% which induces coherent population transfer between the components~\cite{farolfi_2021}. 
Starting from a disordered symmetric phase with $\psi_j=0$ at $t=0$, the system is driven across the 
condensation by quenching the chemical potential according to
\begin{equation}
    \mu(t) = \mu_i + \frac{t}{\tau_Q} (\mu_f - \mu_i),
\end{equation}
where $\tau_Q^{-1}$ sets the quench rate.
% {\color{red}This protocol drives the system through a 
% $\mathbb{Z}_2$ symmetry-breaking transition, leading dynamically to a 
% ferromagnetic phase with finite longitudinal magnetization for $\Omega/\Omega_{\rm cr}<1$~\cite{recati_2021}. {\color{red}write in terms of $\Omega/\mu_f$}}
% Throughout this work we fix $\Omega/\Omega_{\rm cr}=0.36$ (with $\Omega_{\rm cr}=2.77$){\color{red}we can do away with $\Omega_cr$}, placing the 
% system in the ferromagnetic regime. 
For a given choice of $\Omega$, the final chemical 
potential $\mu_f$ is computed accordingly determining the total number of atoms in the steady state 
reached after the quench~\cite{abad_2013}. This protocol facilitates the formation of a weakly coupled Bose--Einstein condensate, characterized by the condition $\Omega/\mu_f \ll 1$. We consider a homogeneous two-dimensional system of area 
$\mathcal{A}=(30\times 30)$ with units of length, temperature, and time given by 
$\sqrt{\hbar^2/(mE_{\rm sc})}$, $E_{\rm sc}/k_B$, and $\hbar/E_{\rm sc}$, respectively. 
% The parameters used are $\mu_i=0.1$, $\mu_f=27.77$, $g=1$, $g_{12}=1.2g$, and $T=10^{-6}$, 
The parameters used are $\mu_i=0.1$, $\mu_f = 27.8$, $g=1$, $g_{12}=1.2g$, and $T=10^{-6}$,
with $\epsilon_{\rm cut}=k_B T\ln2+\mu_f$ fixed to ensure adequate occupation of low-lying modes~\cite{Larcher2018}. 
For convenience, we define the energy scale as $E_{\rm sc}=\mu_{\rm f,p}/27.8$, where $\mu_{\rm f,p}$ is the final chemical potential in physical units (approximately $k_{\rm B}\times0.5 {\rm nK}$ for parameters corresponding to the quasi-two-dimensional BEC experiment in Ref.~\cite{Chomaz2015}). The precise choice of $E_{\rm sc}$ does not influence the physics, since all energies, lengths, and times are rescaled with respect to $E_{\rm sc}$ to render Eq.~\eqref{sgpe} dimensionless. It merely serves as a convenient numerical scale for efficiently solving the SPGPE.
\begin{figure*}[!htbp]
    \centering
    \includegraphics[width=0.85\linewidth]{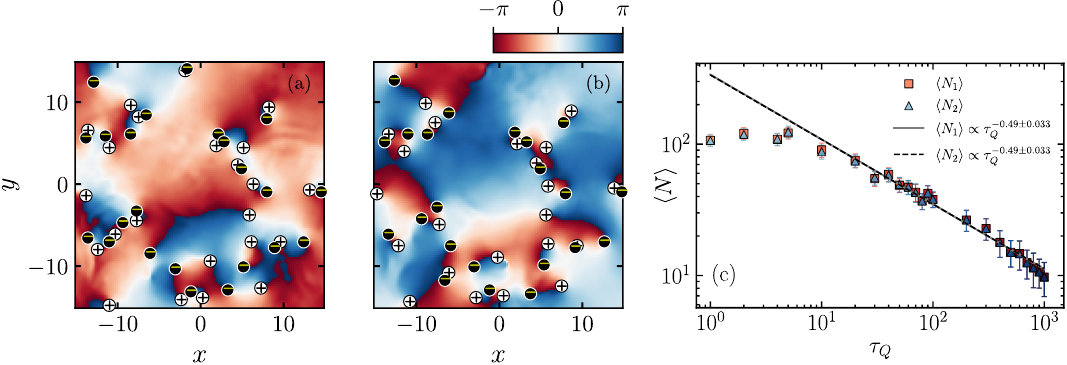}
    \caption{Kibble--Zurek scaling of vortex defects without considering their topological charge evaluated at the equilibration time $t/t_{\rm eq}\sim 1$ for quench times $\tau_Q$ ranging from $1$ to $1000$. 
(a,b) Phase profiles of the two spin components at $t_{\rm eq}$, with vortices and antivortices indicated by positive and negative winding numbers, respectively. 
(c) Universal power-law scaling of the mean vortex number $\langle N \rangle$ with the quench time $\tau_Q$. For the two spin components, the scalings are $\langle N_1 \rangle \propto \tau_Q^{-0.49 \pm 0.033}$ and $\langle N_2 \rangle \propto \tau_Q^{-0.49 \pm 0.033}$. Fits are obtained using a weighted least-squares procedure that accounts for the uncertainty of each data point. Error bars denote one standard deviation over $\mathcal{R}=100$ independent noise realizations.}
    \label{vortex_scaling}
\end{figure*}

\section{Nonequilibrium growth dynamics}
\label{defect_kzm}
\subsection{Equilibration time}
In this section, we analyze condensate formation from vacuum and identify a characteristic dynamical timescale $t_{\rm eq}$ relevant for defect formation. The continuous symmetry-breaking transition is signaled by the growth of the condensate norm
$\mathcal{N}(t)=({1}/{\mathcal{A})\sum_{j=1,2}\int d\mathbf{x}\,
|\psi_j(\mathbf{x},t)|^2}$,
shown in Fig.~\ref{t_eq}(a). During the initial stages of evolution, the condensate begins to develop through the formation of small protodomains with independently chosen phases. However, in this early freeze-out regime (\(t\ll t_{\rm eq}\)), the condensate density remains too low to establish long-range coherence and admit well-formed topological defects~\cite{chesler_2015}.
Following the quench, in practice, the condensate norm initially grows exponentially before crossing over to an adiabatic linear growth regime at the equilibration time \(t_{\rm eq}\), marked by black triangles in Fig.~\ref{t_eq}(a), which defines the characteristic crossover timescale between these two dynamical regimes for finite quench times \(\tau_Q\).
% Following the quench, in practice, the condensate norm exhibits an initial exponential growth before crossing over to an adiabatic linear increase for finite quench times \(\tau_Q\) at the equilibration time \(t_{\rm eq}\), indicated by black triangles in Fig.~\ref{t_eq}(a), as the crossover timescale between these two regimes. 
Around \(t_{\rm eq}\), independently formed protodomains begin to merge, leading to the onset of global phase coherence, spontaneous symmetry breaking, and defect formation. An analogous definition of equilibration time has previously been employed in holographic superfluids~\cite{chesler_2015}. The extraction procedure for \(t_{\rm eq}\) is described in Appendix~\ref{charc_t_eq}.
For \(t<t_{\rm eq}\), the condensate norm for different $\tau_Q$ collapses onto a universal curve when plotted against the scaled time \(t/\sqrt{\tau_Q}\), consistent with the KZ prediction for the freeze-out time \(\hat t\)~\cite{damski_2010, Sonner2015}. The equilibration time thus exhibits the scaling
\begin{equation}
t_{\rm eq} \propto \hat t \propto \tau_Q^{z\nu/(1+z\nu)},
\end{equation}
yielding the exponent \(z\nu/(1+z\nu)=0.47\pm0.002\), in excellent agreement with the KZ prediction for \(z\nu=1\), as shown in Fig.~\ref{t_eq}(b). 
 For fast quenches $\tau_Q \lesssim 5$, we find deviations from the universal scaling, indicating the breakdown of KZM~\cite{chesler_2015, Donadello_2016, Zeng23}.

%\section{Proliferation of quantized circulations and universal dynamics}
\subsection{Universal dynamics of topological defects}

Building on the equilibration time scale $t_{\rm eq}$ identified in the previous section, we now characterize the spontaneous nucleation of vortices. 
% Around $t \sim t_{\rm eq}$, phase-coherent condensate domains with independently chosen phases begin to merge, leading to the nucleation of quantized vortices. 
The vortex core size is set by the healing length, and each vortex is characterized by an integer phase winding number $w$. The quantized circulation around a closed contour enclosing the core is given by~\cite{Pethick_2008}
\begin{equation}
    \oint \mathbf{v}(\mathbf{r}, t)\cdot d\mathbf{l} 
    = \frac{2\pi \hbar}{m} w, \qquad w \in \mathbb{Z},
\end{equation}
where $\mathbf{v}(\mathbf{r},t) \sim \nabla \phi$ is the superfluid velocity and $\phi$ is the condensate phase, defined modulo $2\pi$. Depending on the sign of $w$, the circulation corresponds to a vortex ($w>0$) or an antivortex ($w<0$), as illustrated in Fig.~\ref{vortex_scaling}(a,b). 

We count the total number of vortices (point defects) at $t_{\rm eq}$, irrespective of their topological charge. The vortex number $N$ exhibits a clear power-law dependence on the quench time $\tau_Q$ over the range $\tau_Q = 1$--$1000$. For the two spin components, we find 
\begin{equation}
\langle N_1 \rangle \propto \tau_Q^{-0.49 \pm 0.033}, 
\qquad 
\langle N_2 \rangle \propto \tau_Q^{-0.49 \pm 0.033},
\end{equation}
as shown in Fig.~\ref{vortex_scaling}(c), consistent with the KZ prediction [Eq.~\eqref{defect_KZ}] for point defects in two dimensions ($D=2$, $d=0$). Here, $\langle \cdots \rangle$ denotes averaging over $\mathcal{R}$ independent noise realizations. The fits performed in the scaling regime $\tau_Q = 20$--$1000$ are in quantitative agreement with KZM using mean-field critical exponents $z=2$ and $\nu=1/2$~\cite{mcdonald_15,su_13,Sabbatini_2011,lee_09}. The small deviation from the ideal KZ exponent of $1/2$ can be attributed to uncertainties in the precise determination of $t_{\rm eq}$ and residual finite-size effects.
% \\
% \paolo{3) later we show that considering only FQVs, the scaling improves. is it not that the corrections are trivially because we calculate all vortices and not just FQVs? leading to the conlcusion that indeed one needs to consider F and not H qv for KZM in coupled bose gases?}

%\subsection{Emergence of full quantum vortices}
\begin{figure}[!htbp]
    \centering
    \includegraphics[width=0.9\linewidth]{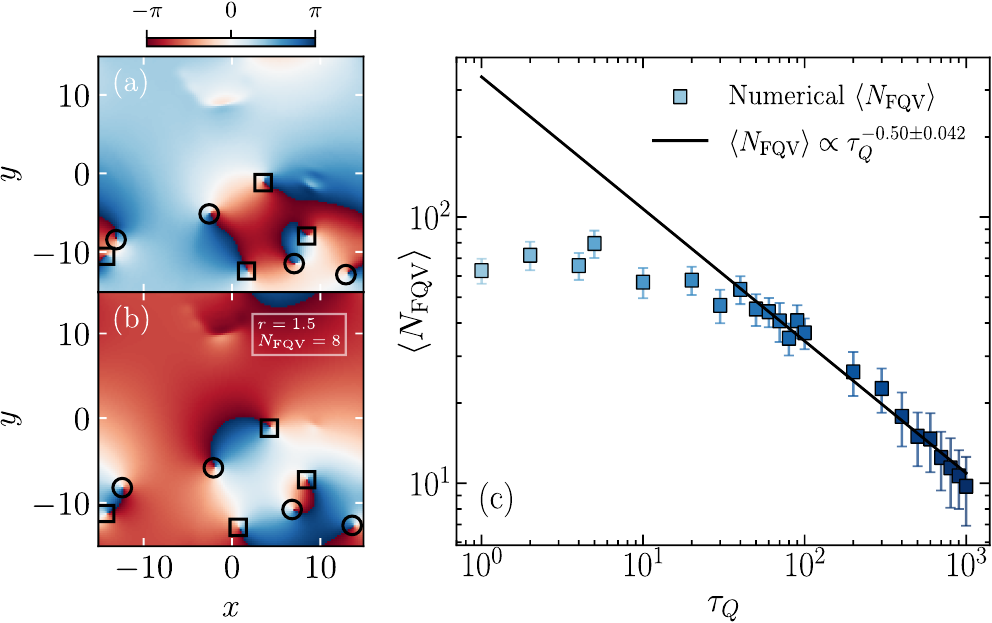}
    \caption{
(a,b) Phase profiles of the two spin components, $\phi_1$ and $\phi_2$, at $t/t_{\rm eq}=1.92$. For representation purpose, a later time snapshot is chosen to clearly highlight vortex pairing, as the number of vortices is relatively small for $\tau_Q=50$. Vortices with winding number $w=-1$ (squares) and $w=+1$ (circles) are indicated in both components. Vortices with equal winding in the two components pair to form full quantum vortices (FQVs), subject to the spatial pairing threshold. 
(c) Mean number of FQVs, $\langle N_{\rm FQV} \rangle$, as a function of the quench time $\tau_Q$ computed at $t/t_{\rm eq }\sim 1$ for the spatial pairing threshold $r=1.5$. The blue markers denote numerical data, with darker shades corresponding to slower quenches. Fits over the range $\tau_Q=40$--$1000$ reveal KZ scaling, $\langle N_{\rm FQV} \rangle \propto \tau_Q^{-0.50 \pm 0.042}$. Fits are performed using a weighted least-squares method that accounts for the statistical uncertainty of each data point. Error bars represent one standard deviation over $\mathcal{R}=100$ stochastic realizations.}
    %\caption{Dependency of the full quantum vortices (FQVs) $N_{\rm FQV}$ at $t_{\rm eq}$ with the quench time $\tau_Q$. Panel (a)-(b) represents the phase profiles of the two spin states $\phi_1$ and $\phi_2$ at $t/t_{\rm eq} = 1.92$ for $\tau_Q = 50$, respectively. The circulations having a winding number $w = -1$ are indicated with the square markers, and the winding number $w = +1$ are marked with the circular markers in both the phases. Vortices with equal winding from both the phases pair up to form a full quantum vortex (FQV) subject to the spatial threshold. Panel (c) represents the dependency of the FQVs with the quench time $\tau_Q$. The blue markers are the numerical $\langle N_{\rm FQV}\rangle$ with increasing darker shed for increasing quench rate. Fitting of the numerical data ranging from $\tau_Q = 30$ to $\tau_Q = 1000$ reveals a KZM like scaling: $\langle N_{\rm FQV} \rangle \propto \tau_Q^{-0.50 \pm 0.014}$. The error bars show one standard deviations across $\mathcal{R} = 100$ noise realizations.} 
    \label{full-vortex-scaling}
\end{figure}

In addition, we observe that vortices with the same (signed) winding number in the two spin components bind to form a composite defect, subject to a finite spatial resolution, referred to as a full quantum vortex (FQV)~\cite{Dominici_2015, Dagvadorj_2023}. Representative FQVs are shown in Fig.~\ref{full-vortex-scaling}(a,b), with different markers indicating different (signed) winding numbers. To identify FQVs, we introduce a spatial pairing threshold $r=1.5$ in units of $\sqrt{\hbar^2/(mE_{\rm sc})}$. Unless otherwise mentioned, we use this value consistently throughout this work as it avoids underestimating the number of identified FQVs while preserving the expected KZ scaling behavior across the range of $\tau_Q$ investigated.

We count the number of FQVs at the equilibration time $t_{\rm eq}$ for quenches in the range $\tau_Q=1$--$1000$. A least-squares fit yields a universal power-law scaling, $\langle N_{\rm FQV} \rangle \sim \tau_Q^{-(D-d)\nu/(1+z\nu)}$ with an exponent $(D-d)\nu/(1+z\nu)=0.50 \pm 0.042$, in excellent agreement with the KZ prediction for point defects ($d=0$), as shown in Fig.~\ref{full-vortex-scaling}(c). 
These demonstrate that in coherently coupled Bose mixtures, FQVs emerge during the dynamics of continuous symmetry-breaking transition via KZM and the subsequent phase-ordering dynamics.

We further test the robustness of this identification by varying the resolution threshold over the range $r=\{0.5,\,0.75,\,1,\,1.5,\,2\}$ when identifying vortices in the complementary phase fields $\phi_1$ and $\phi_2$. The emergence of FQVs is attributed to the fact that $\Omega > (g_{12}-g){\mathcal N}(t)/2$ at $t/t_{\rm eq}=1$, under which the separation between the constituent vortices decreases ~\cite{kasamatsu2004,son_02}.
% A larger $\Omega$ enhances the tension of the domain wall connecting the constituent vortices reducing their separation
For fast quench rates, the threshold plays a crucial role, leading to larger relative separations between vortices in the two components. Consequently, a significant fraction of vortices fail to bind into composite defects, giving rise to the coexistence of full-quantum vortices (FQVs) and half-quantum vortices (HQVs). The spatial statistics and stochastic geometry analyses presented in this work are therefore focused on the quench time $\tau_Q=50$, which lies in this intermediate regime where a substantial population of FQVs coexists with a smaller but non-negligible number of HQVs. This regime is particularly valuable because it allows us to directly compare the spatial organization of vortices in the individual condensate components with that of the composite FQVs. By contrast, for slower quenches, vortex pairing becomes nearly complete and almost all vortices form composite defects, leaving little distinction between the statistics of the individual components and those of the FQVs. Throughout this work, the position of an FQV is defined as the center of mass of its constituent vortices, namely the midpoint of the line segment connecting the two bound vortex cores. Further details are provided in the Appendix~\ref{fqv_threshold}. 

\section{Universal spatial statistics and stochastic geometry: beyond kzm}
\label{Stochastic_geometry}
\subsection{Defect number statistics} 
  According to the geodesic rule~\cite{kibble_1980}, the merging of phase domains produces circulating loops with a net $2\pi$ phase winding with a certain success probability $p$. Recent studies have shown that full counting statistics of topological defects reveal universal behavior beyond the scope of mean-field KZ scaling~\cite{Ruiz_2020, Thudiyangal_2024}. In this framework, defect formation events are treated as independent and uniformly distributed. The total number of statistically independent possible defect-formation sites is set by the KZ correlation length $\hat{\xi}$,
\begin{equation}
    \mathcal{N}_d = \frac{\mathcal{A}}{f \hat{\xi}^2},
\end{equation}
where $f$ is a fudge factor accounting for the average number of domains required to generate a quantized circulation~\cite{Ruiz_2020}. The number of successful events $N$ out of $\mathcal{N}_d$ Bernoulli trials follows a binomial distribution,
\begin{equation}
        P(N) = \binom{\mathcal{N}_d}{N} p^N (1 - p)^{\mathcal{N}_d - N},
        \label{binomial}
\end{equation}
centered around the KZ mean defect number
\begin{equation}
\langle N \rangle = p\,\mathcal{N}_d 
\sim \frac{p\mathcal{A}}{\hat{\xi}^D} 
\sim \tau_Q^{-(D-d)\nu/(1+z\nu)},
\end{equation}
which, for point defects ($d=0$) with mean-field exponents $z=2$ and $\nu=1/2$, is verified in Fig.~\ref{vortex_scaling}(c). 

In the limit of large $\mathcal{N}_d$, the binomial distribution approaches a normal distribution by the central limit theorem given by~\cite{Thudiyangal_2024},
\begin{equation}
    P(N) = \frac{1}{\sqrt{2\pi (1 - p) \langle N \rangle}} 
\exp\!\left(
    -\frac{(N - \langle N \rangle)^2}{2(1 - p) \langle N \rangle}
\right),
\label{normal}
\end{equation}
with the success probability related to the variance as~\cite{Ruiz_2020}
\begin{equation}
p = 1 - \frac{\mathrm{Var}(N)}{\langle N \rangle}.
\end{equation}

\begin{figure}[!htbp]
    \centering
    \includegraphics[width=0.9\linewidth]{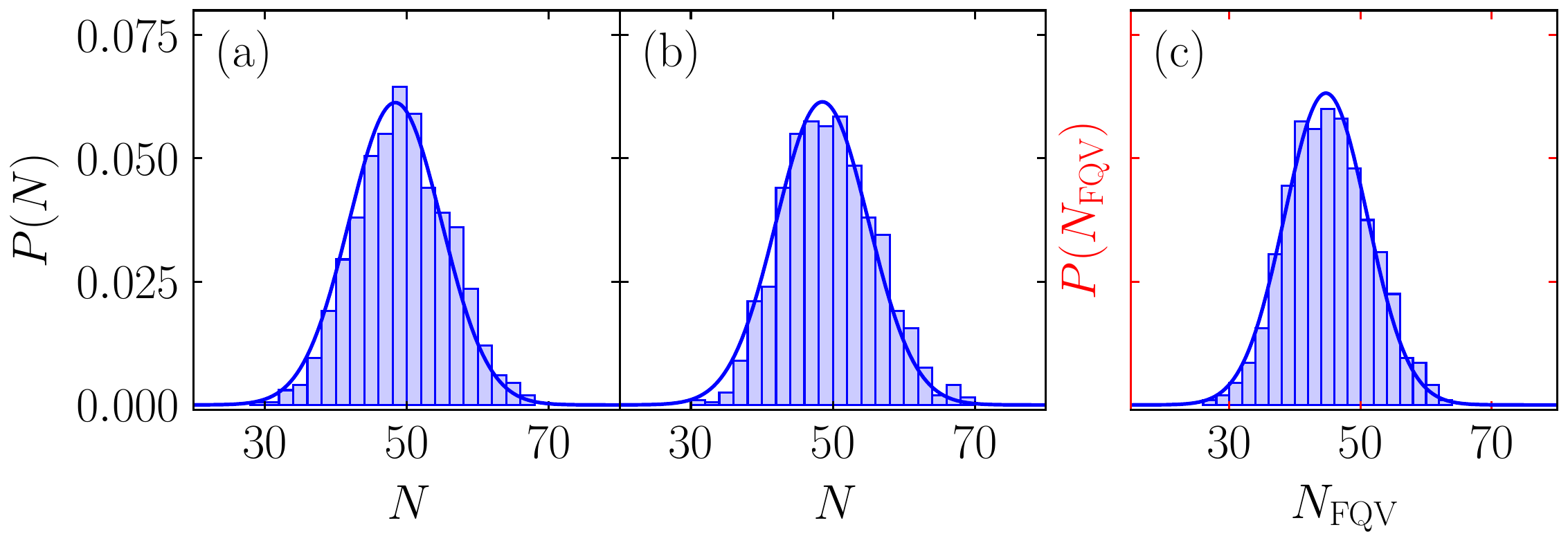}
    \caption{Panels (a)-(b) show the probability distribution of the number of point defects (without considering topological charge) at $t/t_{\rm eq}=1$ for the two spin components, respectively for the quench time $\tau_Q = 50$. The solid curves represent the theoretical normal distribution given by Eq.~\eqref{normal}. Panel (c) displays the corresponding distribution for FQVs, with the solid line again indicating Eq.~\eqref{normal}. All histograms are constructed from $\mathcal{R}=1000$ independent noise realizations. }
    \label{number_dist}
\end{figure}
We compute the full distribution of quantized circulations $N$ in both spin components [Fig.~\ref{number_dist}(a,b)] and observe a clear Gaussian profile in excellent agreement with Eq.~\eqref{normal}. We further evaluate the number distribution of full quantum vortices $N_{\rm FQV}$, identified via the spatial threshold criterion $r = 1.5$, and find consistency with the underlying binomial statistics [Fig.~\ref{number_dist}(c)] through Eq.~\eqref{normal}.

% We calculate the number distribution of the quantized circulations $N$ in both the spin states shown in the Fig.~\ref{number_dist}(c)-(d) and observe a gaussian signature which is in good agreement with the theoretical distribution given in the Eq.~\eqref{normal}. We also calculate the number distribution for the FQVs $N_{\rm FQV}$ subject to the spatial threshold and characterize the underlying Binomial distribution shown in the Fig.~\ref{number_dist}(e). 

\subsection{Universal spatial statistics via Poisson point process}
\label{sub:unconditioned_dist}
\subsubsection{Pairwise distance distribution}
\label{subsubsec:poissonPP}

We treat the vortex positions (without distinguishing topological charge) at $t \sim t_{\rm eq}$ as statistically independent events and model them using a homogeneous PPP, with the mean defect number $\langle N \rangle$ governed by KZ scaling~\cite{Thudiyangal_2024,campo_2022}. To probe the underlying spatial correlations of spontaneously generated vortices in both spin components, we analyze the pairwise vortex--vortex distance distribution.

\begin{figure}[!htbp]
    \centering
    \includegraphics[width=0.9\linewidth]{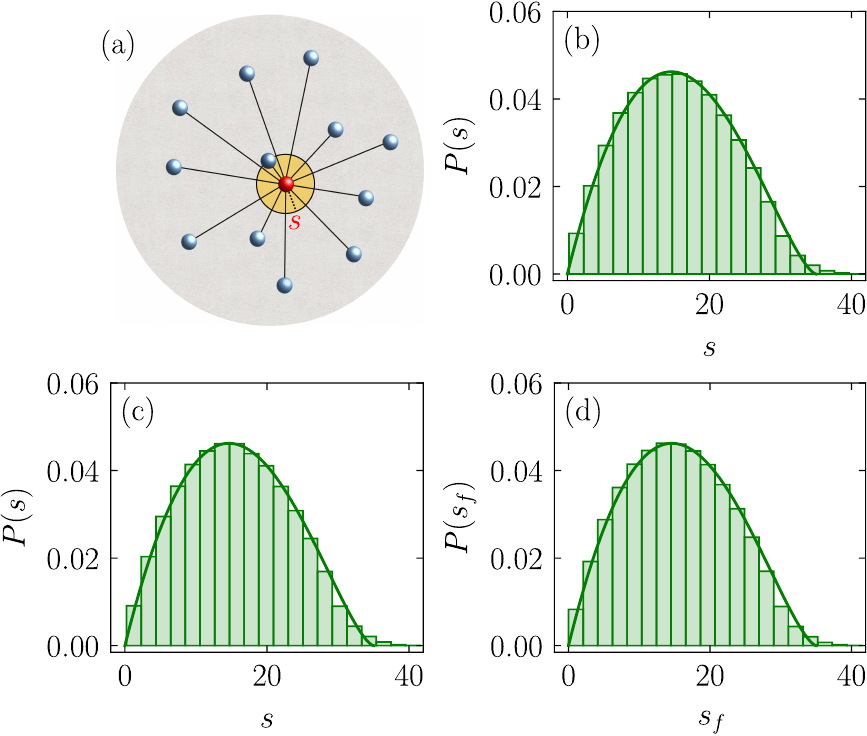}
    \caption{Pairwise distance distribution of vortices at $t/t_{\rm eq}=1$ for the quench time $\tau_Q=50$. Panel (a) corresponds to a schematic representation of random points (vortices) with a reference vortex (red) on a circular disk of radius $R$. The $1^{\rm st}$ nearest neighbor (NN) of the reference vortex is indicated using a circle with radius $s$. (b)-(c) Distance distributions of vortices for the two spin components ($j=1,2$), irrespective of their topological charge. (d) Pairwise distance distribution of FQVs. 
All histograms are constructed from $\mathcal{R}=1000$ stochastic realizations using $20$ bins. 
The solid curves represent the theoretical disk line-picking distribution given by Eq.~\eqref{dist_disk} for a disk of radius $R=17.5$. }
    \label{ppp_dist}
\end{figure}

For vortices located at positions $\mathbf{r}$ and $\mathbf{r}'$, we compute their separation $s = |\mathbf{r}-\mathbf{r}'|$. Taking each vortex in turn as a reference [see Fig.~\ref{ppp_dist}(a)], we calculate distances to all other vortices and aggregate over all pairs and independent noise realizations. If vortices are uniformly and independently distributed over a disk of radius $R$, the pairwise distance distribution reduces to the classical disk line-picking problem in geometric probability~\cite{Mathai1999}. The corresponding probability density is
\begin{equation}
P(s) = \frac{4s}{\pi R^2}\left[\arccos\left(\frac{s}{2R}\right) 
- \frac{s}{2R}\sqrt{1-\frac{s^2}{4R^2}}\right].
\label{dist_disk}
\end{equation}

We construct the distance distribution at $t/t_{\rm eq}=1$ by evaluating all pairwise separations of stochastically generated vortices across different realizations. The resulting distributions for both spin components [Fig.~\ref{ppp_dist}(b,c)], show excellent agreement with Eq.~\eqref{dist_disk}, supporting the PPP description. 
We perform the same analysis for full quantum vortices (FQVs), aggregating pairwise distances $s_f$ across realizations [Fig.~\ref{ppp_dist}(d)]. The FQV distribution likewise matches the theoretical prediction of Eq.~\eqref{dist_disk}, indicating that composite defects also follow PPP statistics. Thus, although vortex positions vary randomly between realizations during growth into the symmetry-broken phase, the aggregated statistics reveal an emergent universal geometric structure consistent with a homogeneous PPP governed by KZM-determined defect density.
% We also calculate pairwise distances for FQVs and construct the distribution from different noise realizations shown in the Fig.~\ref{ppp_dist}(c). It is evident that FQVs follow PPP and matches well with the theoretical distribution given in the Eq.~\eqref{dist_disk}. 
% {\color{red}fig 5d might appear first. sequential description is hampered. also "r" is already defined. we might have to change either of it.}

\subsubsection{$n^{\rm th}$---order spacing statistics}
\begin{figure*}
    \centering
    \includegraphics[width=0.9\linewidth]{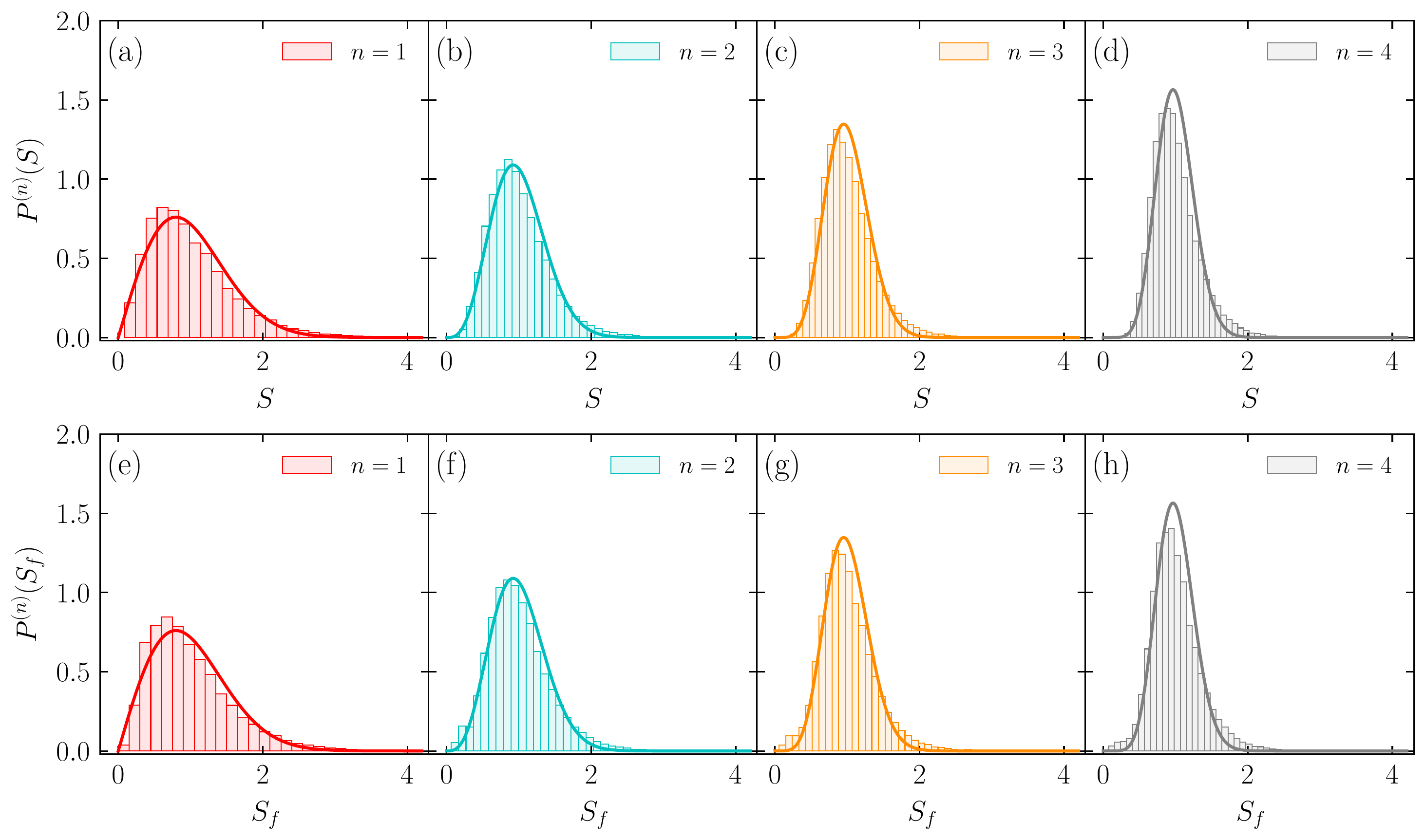}
    \caption{Nearest-neighbor (NN) spacing distributions at $t/t_{\rm eq}=1$ for the quench time $\tau_Q=50$. 
(a)--(d) NN spacing distribution $P^{(n)}(S)$ of quantized vortices in spin component $j=1$ for different NN orders. 
(a) First NN ($n=1$), where the solid curve represents the Wigner--Dyson distribution given by Eq.~\eqref{wd}. 
(b)--(d) Higher-order NN spacing distributions ($n=2,3,4$), with solid curves corresponding to the theoretical prediction for a 2D Poisson point process given by Eq.~\eqref{nth-ND}. 
(e)--(h) NN spacing distributions $P^{(n)}(S_f)$ for full quantum vortices (FQVs) for NN orders $n=1,2,3,4$, respectively. 
All histograms are constructed from $\mathcal{R}=1000$ stochastic realizations using $40$ bins. }
    \label{nth-order_dist}
\end{figure*}

We further investigate the spatial statistics of defect positions by analyzing the spacing between vortices and their nearest neighbors (NN)~\cite{campo_2022,Thudiyangal_2024}. This analysis can be generalized to the $n^{\rm th}$---order spacing by selecting a vortex as a reference and determining the probability that the next defect lies outside a circle of radius $s$ centered at the reference vortex, as illustrated schematically in Fig.~\ref{ppp_dist}(a) for the first NN ($n=1$).

For a two-dimensional PPP, the $n^{\rm th}$---order spacing distribution is given by~\cite{haake2010,Sakhr_2006}
\begin{equation}
P^{(n)}(S) = P(S;n,2) = \frac{2 \alpha^n}{\Gamma(n)} S^{2n-1} e^{-\alpha S^2},
\label{nth-ND}
\end{equation}
where $\alpha = [\Gamma(n+\tfrac{1}{2})/\Gamma(n)]^2$ and $S = s/\langle s \rangle$ is the normalized spacing. For $n=1$, this expression reduces to the 
% {\color{red}first?} nearest-neighbor spacing distribution, which coincides with 
Wigner--Dyson (WD) distribution~\cite{Guhr1998,Sakhr_2006}:
\begin{equation}
P(S) = \frac{\pi}{2} S \exp\left(-\frac{\pi}{4} S^2\right).
\label{wd}
\end{equation}

We compute the spacing distributions for vortices in both spin components without distinguishing their topological charge, irrespective of whether vortices from complementary phases pair to form FQVs. The first nearest-neighbor distribution is shown in Fig.~\ref{nth-order_dist}(a), where the solid line represents the theoretical WD distribution. The second-, third-, and fourth-nearest-neighbor spacing statistics are presented in Fig.~\ref{nth-order_dist}(b)--(d) for $n=2,3,4$, respectively. In each case, the numerical distributions obtained from multiple stochastic realizations follow the theoretical predictions of Eq.~\eqref{nth-ND}, confirming the PPP description of the defect spatial statistics. 
Furthermore we analyze the spatial statistics of FQV
for the quench time $\tau_Q=50$ by computing the $n^{\rm th}$---order nearest-neighbor (NN) spacing distribution. The distributions $P^{(n)}(S_f)$ are constructed by aggregating NN distances of FQVs from $\mathcal{R}$ independent stochastic realizations. Figs.~\ref{nth-order_dist}(e)--(h) show the resulting distributions for the first to fourth NN spacings, corresponding to $n=1,2,3,4$, respectively. Here $S_f=s_f/\langle s_f\rangle$ denotes the normalized spacing between FQVs. We find that the numerically obtained spacing statistics of FQVs closely follow the analytical prediction for the $n^{\rm th}$-order spacing distribution of a two-dimensional PPP given by Eq.~\eqref{nth-ND}. 
%{\color{red}comment on deviation?like earlier}
% Neglecting the HQVs for the quench time $\tau_Q=50$, we calculate the $k^{\rm th}$\textemdash order NN of FQVs and construct the distribution $P^{(k)}(S_f) = P^{(k)}(S)$ by aggregating all the NN distances depending on the NN order $(k)$ from $\mathcal{R}$ different random trajectories shown in the Fig.~\ref{nth-order_dist}(e)-(h) for the $1^{\rm st}(k=1)$, $2^{\rm nd}(k=2)$, $3^{\rm rd}(k=3)$, and $4^{\rm th}(k=4)$\textemdash NN spacing respectively. Here $S_f$ is the mean spacing of FQVs. We observe that the spatial statistics of FQVs calculated numerically also follow the analytically estimated probability distribution given in the Eq.~\eqref{nth-ND} for the $k^{\rm th}$\textemdash NN corresponding to the 2D PPP. 

\begin{figure}[!htbp]
    \centering
    \includegraphics[width=0.9\linewidth]{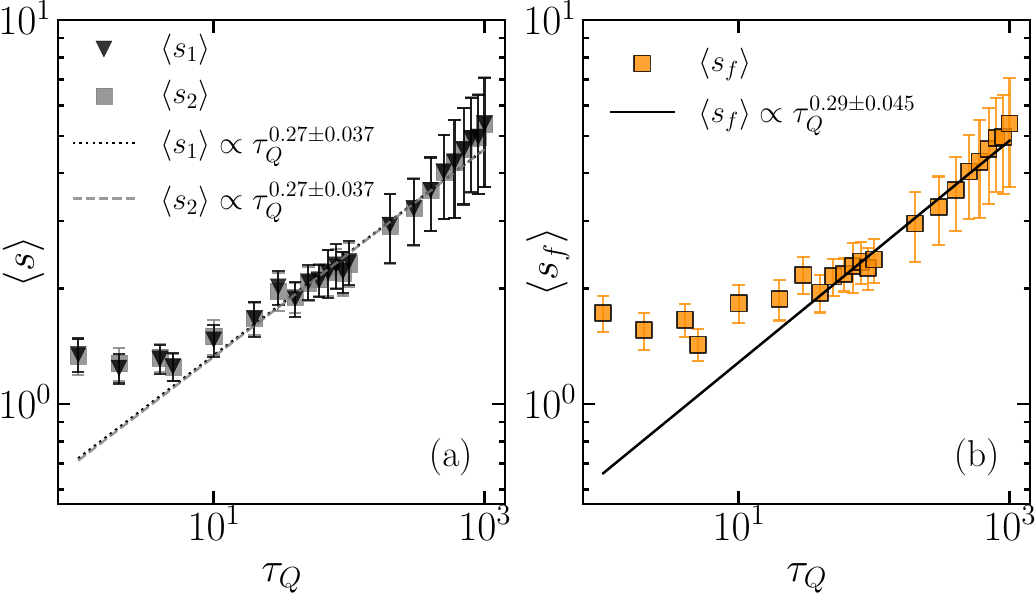}
    \caption{ 
(a) Mean spacing of quantized vortices  $\langle s \rangle$ at the equilibration time $t=t_{\rm eq}$ computed independently for the two spin components as a function of the quench time $\tau_Q$, denoted by $\langle s_1 \rangle$ and $\langle s_2 \rangle$, for quenches in the range $\tau_Q=1$--$1000$. Both follow Kibble--Zurek scaling, $\langle s_1 \rangle \propto \tau_Q^{0.27 \pm 0.037}$ and $\langle s_2 \rangle \propto \tau_Q^{0.27 \pm 0.037}$. 
(b) Mean spacing of FQVs, $\langle s_f \rangle$, showing KZM-like scaling $\langle s_f \rangle \propto \tau_Q^{0.29 \pm 0.045}$ obtained from fits in the range $\tau_Q=40$--$1000$. 
Fits are performed using a weighted least-squares method accounting for the uncertainty of each data point. Error bars denote one standard deviation over $\mathcal{R}=100$ stochastic realizations.}
    \label{spacing_scaling}
\end{figure}

% {\color{red} should this part come at the beginning? do you mean to say that to calculate S, you require $<s>$ which obeys scaling law?, or we can write s is an input which is computed like this?}

For $n=1$, in Eq.~\eqref{nth-ND} the mean nearest-neighbor spacing follows
\begin{equation}
    \langle s \rangle = \frac{\sqrt{\pi}}{2}\,\hat{\xi}
    \propto \tau_Q^{\frac{\nu}{1+z\nu}},
    \label{spacing_kzm}
\end{equation}
where $z$ and $\nu$ are the mean-field critical exponents as stated earlier in the text. 
We first evaluate the mean spacing of quantized vortices at $t \sim t_{\rm eq}$, treating vortices in the two spin components independently and ignoring their topological charge. For quenches in the range $\tau_Q=1$--$1000$, we obtain the scalings $\langle s_1 \rangle \propto \tau_Q^{0.27 \pm 0.037}$ and $\langle s_2 \rangle \propto \tau_Q^{0.27 \pm 0.037}$ for the two components, respectively. These results are in good agreement with the KZM prediction of Eq.~\eqref{spacing_kzm}, as shown in Fig.~\ref{spacing_scaling}(a).
Next, we analyze the spacing statistics of the FQVs, obtained by pairing vortices from the complementary spin components. 
% The mean FQV spacing $\langle s_f \rangle$ is computed without distinguishing whether the pairs correspond to $(+,+)$ or $(-,-)$ circulations. 
The dependence of $\langle s_f \rangle$ on the quench time $\tau_Q$, shown in Fig.~\ref{spacing_scaling}(b), yields a power-law scaling with exponent $\nu/(1+z\nu)=0.29 \pm 0.045$. This numerical result is consistent with the theoretical KZ scaling given in Eq.~\eqref{spacing_kzm}. 

\begin{figure*}[!htbp]
    \centering
    \includegraphics[width=0.9\linewidth]{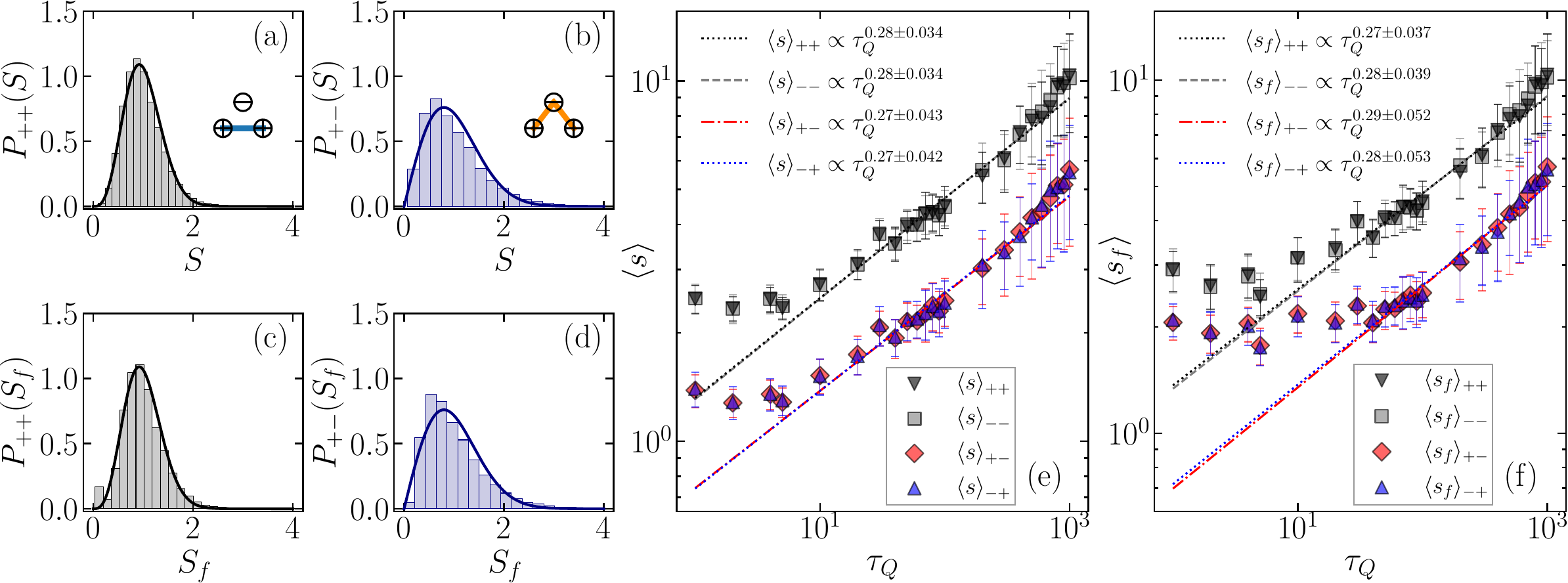}
    \caption{Universality in spacing statistics conditioned on vortex topological charge at $t/t_{\rm eq}=1$ for $\tau_Q=50$. 
(a)-(b) First nearest-neighbor (NN) spacing distributions of vortices in the $j=1$ component conditioned on charge: (a) $P_{++}(S)$ for $\oplus$--$\oplus$ vortices and (b) $P_{+-}(S)$ for $\oplus$--$\ominus$ vortices. Solid curves correspond to the theoretical predictions of Eq.~\eqref{nth-ND} with $n=2$ and $n=1$ (Wigner--Dyson), respectively. 
(c)-(d) First NN spacing distributions of FQVs formed by pairing two vortices from complementary phases: (c) $P_{++}(S_f)$ and (d) $P_{+-}(S_f)$, with solid curves given by Eq.~\eqref{nth-ND} for $n=2$ and $n=1$, respectively. 
(e) Kibble--Zurek scaling of the conditioned mean spacing $s_{cc'}$ of vortices for both spin components. 
(f) KZ scaling of the mean spacing of FQVs. 
Histograms are constructed from $\mathcal{R}=1000$ stochastic realizations (25 bins). Error bars represent one standard deviation over $\mathcal{R}=100$ realizations, and fits are obtained using weighted least squares.}
    \label{spacing_winding_kzm}
\end{figure*}

\subsubsection{Universal statistics of vortices conditioned on the topological charge}
% In the sec.~\ref{sub:unconditioned_dist}, we model the location of vortices using a PPP model which inherently assumes every vortex on equal footing, without considering their topological charges, and any logarithmic pair potential between them~\cite{Campbell_1987}. 
We further analyze the spatial statistics of vortices by considering their topological charges~\cite{Campo2021}.
Around $t\sim t_{\rm eq}$, vortices become clearly visible as density-depleted cores. Each vortex is characterized by a topological charge $c^{\pm}$, corresponding to a phase winding number $w=\pm 1$. The presence of topological charge introduces effective correlations between vortices: vortices with identical charge repel each other, while oppositely charged vortices exhibit attraction.

To quantify these correlations, we consider a reference vortex with charge $c$ and compute the probability $P_{cc'}(S)\,dS$ of finding the nearest vortex with charge $c'$ at a distance between $S$ and $S+dS$, while ignoring vortices with the same charge as the reference vortex inside the circular region of radius $S$. The corresponding nearest-neighbor spacing distributions for different charge combinations are summarized in Table~\ref{spacing_prob_table}.

\begin{table}[!h]
    \centering
    \begin{tabular}{|c|c|c|c|c|}
        \hline
        Topological charge & $c^+c^+$ & $c^-c^-$ & $c^+c^-$ & $c^-c^+$  \\ \hline
        Probability distribution & $P_{++}(S)$ & $P_{--}(S)$ & $P_{+-}(S)$ & $P_{-+}(S)$ \\
        \hline
    \end{tabular}
    \caption{Nearest-neighbor spacing distributions for vortices with different combinations of topological charge.}
    \label{spacing_prob_table}
\end{table}

We compute the distributions $P_{cc'}(S)$ by aggregating nearest-neighbor distances from $\mathcal{R}$ independent stochastic realizations. Figure~\ref{spacing_winding_kzm}(a) shows the distribution $P_{++}(S)$ for positively charged vortices in the $j=1$ spin component. The numerical data agree well with the theoretical prediction
\[
P_{cc}(S)=2r^4 S^3 e^{-r^2S^2},
\]
with $r=3\sqrt{\pi}/4$~\cite{Thudiyangal_2024}, which corresponds to Eq.~\eqref{nth-ND} with $n=2$ and $\alpha=r^2$. This behavior reflects the effective repulsion between vortices with identical charge, leading to a larger characteristic separation. By symmetry, the same behavior is expected for negatively charged vortices, i.e., $P_{++}(S)=P_{--}(S)$.

We also compute the distribution $P_{+-}(S)$ corresponding to the nearest-neighbor spacing between vortices with opposite charges, shown in Fig.~\ref{spacing_winding_kzm}(b). In this case, the numerical distribution follows the theoretical prediction of Eq.~\eqref{nth-ND} with $n=1$, which reduces to the Wigner--Dyson form. This reflects the effective attraction between oppositely charged vortices: when a vortex with charge $c^+$ is chosen as the reference, the nearest neighbor is most likely a vortex with charge $c^-$. By symmetry, we expect $P_{+-}(S)=P_{-+}(S)$. The distributions $P_{--}(S)$ and $P_{-+}(S)$, as well as the corresponding results for the $j=2$ spin component, are presented in the Appendix~\ref{spacing_appendix}.
We now extend the analysis to FQVs. 
% which act as composite topological defects formed by pairing vortices from the complementary phases. 
An FQV is assigned a positive (negative) charge when two positively (negatively) charged vortices, $c^+$ ($c^-$), from the two components pair together.
Using this definition, we compute the first nearest-neighbor (NN) spacing distributions for FQVs. The numerical distributions $P_{++}(S_f)$ and $P_{+-}(S_f)$ are constructed from histograms of NN distances across $\mathcal{R}$ stochastic realizations, as shown in Fig.~\ref{spacing_winding_kzm}(c) and Fig.~\ref{spacing_winding_kzm}(d), corresponding to equally charged and oppositely charged FQVs, respectively.
Figure~\ref{spacing_winding_kzm}(c) shows the distribution $P_{++}(S_f)$ obtained by considering only positively charged FQVs. The numerical distribution follows the theoretical prediction of Eq.~\eqref{nth-ND} with $n=2$, shown by the solid curve. This behavior reflects the effective repulsion between equally charged FQVs, which increases their characteristic spacing.
In contrast, Fig.~\ref{spacing_winding_kzm}(d) shows the nearest-neighbor distribution $P_{+-}(S_f)$ for positively charged FQVs when only negatively charged FQVs are considered as neighbors. In this case, the numerical distribution closely follows the Wigner--Dyson form given by Eq.~\eqref{wd}, shown by the solid line. 
%This behavior is consistent with the effective attraction between oppositely charged FQVs.
% Thus far, we calculate the distributions of the vortices independently from both the phases. We further study the distribution of FQVs by considering as a composite defect with topological winding. A FQV is considered to be positively charged when two positively charged $c^+$ vortices from the complementary phases pair up and vice versa. We similarly calculate the $1^{\rm st}$\textemdash NN spacing and construct the numerical distributions $P_{++}(S_f)$ and $P_{+-}(S_f)$ using histograms shown in the Fig.~\ref{spacing_winding_kzm}(c) and \ref{spacing_winding_kzm}(d) for positive-positive and positive-negative FQVs, respectively. In Fig.~\ref{spacing_winding_kzm}(c), we calculate the first NN of each FQVs and find to satisfy the theoretical distribution given in the Eq.~\eqref{nth-ND} with $k = 2$ shown in the Fig.~\ref{spacing_winding_kzm}(c) with the solid line. This is due to the repulsion between the similarly charged FQVs and there exist a oppositely charged FQV within the distance $0$ (reference FQV with $+$ winding) to $S$ (location from the reference FQV with same winding). In Fig.~\ref{spacing_winding_kzm}(d), we estimate the NN spacing distribution of FQVs of positive winding only considering the FQVs with negative winding. The $1^{\rm st}$\textemdash NN spacing distribution $P_{+-}(S_f)$ is in good agreement with the theoretical WD distribution given is the Eq.~\eqref{wd} indicated by the solid line. 

In the conditioned spacing statistics, the KZ universality appears through the power-law scaling of the mean nearest-neighbor distances $s_{cc'}$:
\begin{equation}
    \langle s \rangle_{cc'} \sim \hat \xi \propto \tau_{Q}^{\frac{\nu}{1 + z\nu}},
    \label{conditioned_spacing_kz}
\end{equation}
where $c,c'=\pm$ denote the topological charges of the vortices. Accordingly, the conditioned mean spacings are denoted by $\langle s \rangle_{++}$, $\langle s \rangle_{--}$, $\langle s \rangle_{+-}$, and $\langle s \rangle_{-+}$. 

For vortices with identical charges, the corresponding probability distributions satisfy $P_{++}(S)=P_{--}(S)$, yielding 
%proportioanlity becoming equals?
\[
\langle s \rangle_{++} = \langle s \rangle_{--} = \frac{3}{2}\sqrt{\pi}\,\langle s \rangle,
\]
where $\langle s \rangle$ is the mean spacing of the unconditioned distribution given in Eq.~\eqref{spacing_kzm}. We compute $\langle s \rangle_{++}$ and $\langle s \rangle_{--}$ from the vortex configurations of individual spin components (shown here for $j=1$) over quench times $\tau_Q=1$--$1000$. The resulting scalings,
\[
\langle s \rangle_{++} \propto \tau_Q^{0.28 \pm 0.034}, 
\qquad
\langle s \rangle_{--} \propto \tau_Q^{0.28 \pm 0.034},
\]
are consistent with the KZ prediction.
 
%Error bars represent one standard deviation over multiple stochastic realizations.
% In conditioned spacing distribution, the KZ universality is hidden in the power law scaling of the mean nearest-neighbor distances $s_{cc'}$:
% \begin{equation}
%     \langle s \rangle_{cc'} \sim \hat \xi \propto \tau_{Q}^{\frac{\nu}{1 + z\nu}}.
%     \label{conditioned_spacing_kz}
% \end{equation}
% Depending on the charge combination, mean NN spacing is denoted as $\langle s \rangle_{++}$, $\langle s \rangle_{--}$, $\langle s \rangle_{+-}$, and $\langle s \rangle_{-+}$. The mean spacing corresponds to the probability distribution $P_{++}(S) = P_{--}(S)$ is given by $\langle s \rangle_{++} = \langle s \rangle_{--} = (3/2)\sqrt{\pi} \langle s \rangle$. The mean distance $\langle s \rangle$ is the mean spacing for unconditioned spacing distribution given in the Eq.~\eqref{spacing_kzm}. We calculate $\langle s \rangle_{++}$ and $\langle s \rangle_{--}$ of quantized vortices {\color{red}from both the phases why both? isnt for j=1?} independently for different quench times ranging from $\tau_Q = 1$ to $\tau_Q = 1000$ and find the universal KZ scaling as: $\langle s \rangle_{++} \propto \tau_Q^{0.29 \pm 0.023}$ and $\langle s \rangle_{--} \propto \tau_Q^{0.29 \pm 0.024}$. 
\begin{figure}[!htbp]
    \centering
    \includegraphics[width=0.9\linewidth]{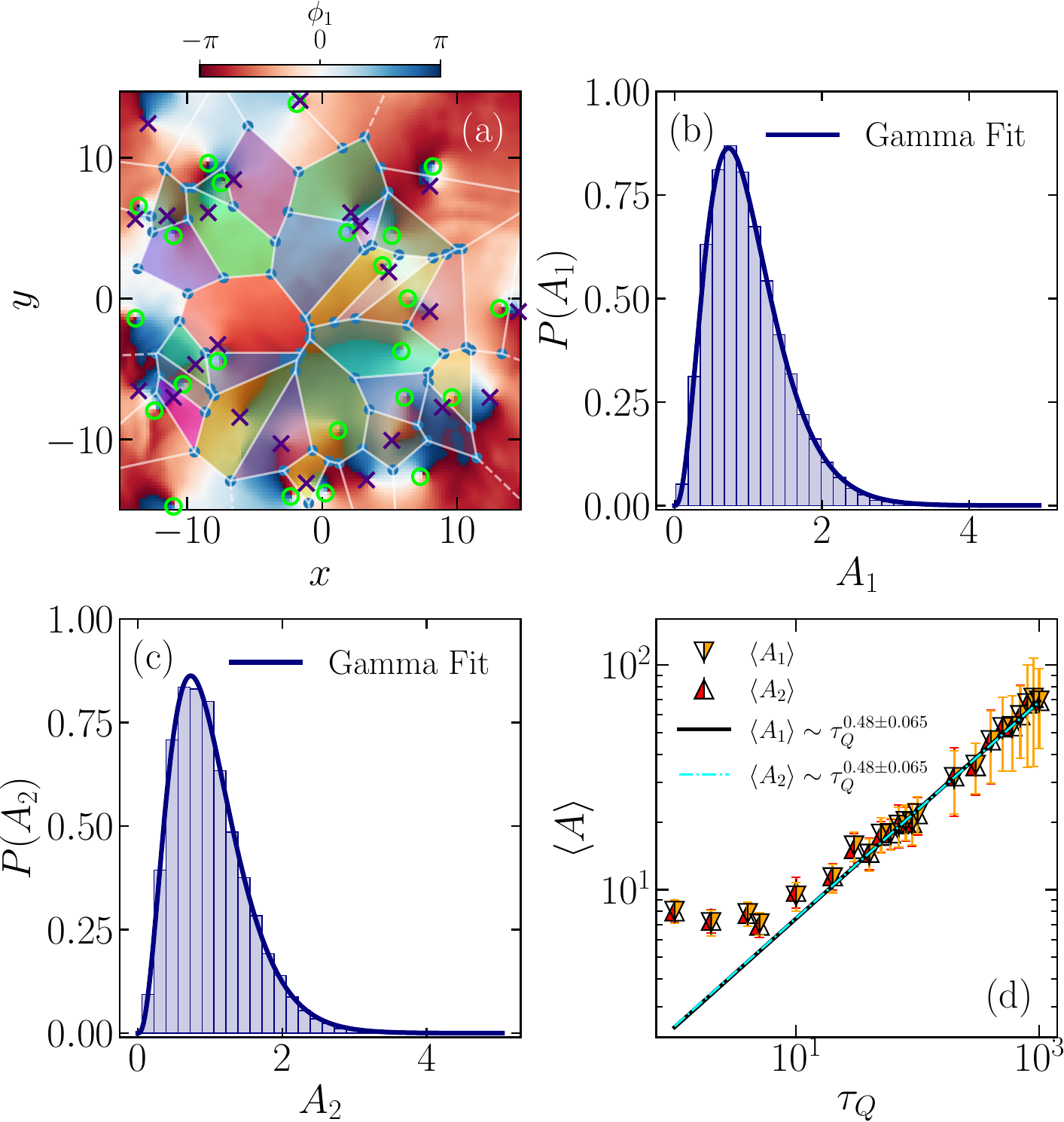}
    \caption{Universal stochastic geometry from Voronoi cell-area statistics at $t \sim t_{\rm eq}$ for $\tau_Q=50$. 
(a) Voronoi tessellation of the phase field for the spin component $j=1$. Colored cells lie within the simulation domain $[-15,15]$. Circular (cross) markers denote vortices with $+2\pi$ ($-2\pi$) phase circulation. 
(b)-(c) Voronoi cell-area distributions $P(A_1)$ and $P(A_2)$ for vortices in the $j=1$ and $j=2$ components, respectively. Histograms (30 bins) are constructed from $\mathcal{R}=500$ stochastic realizations. The solid curve shows the gamma distribution of Eq.~\eqref{gamma} with fitting parameters as mentioned in text. 
(d) Mean Voronoi cell area versus quench time $\tau_Q$. Fits reveal Kibble--Zurek scaling $\langle A_1\rangle \propto \tau_Q^{0.48\pm0.065}$ and $\langle A_2\rangle \propto \tau_Q^{0.48\pm0.065}$ consistent with Eq.~\eqref{voronoi_pl}. Error bars denote one standard deviation over $\mathcal{R}=100$ realizations; fits use weighted least squares.}
    \label{voronoi_panel}
\end{figure}
For vortices with opposite charges, the distributions satisfy $P_{+-}(S)=P_{-+}(S)$, leading to
\[
\langle s \rangle_{+-} = \langle s \rangle_{-+} = \sqrt{2}\,\langle s \rangle.
\]
By computing $\langle s \rangle_{+-}$ and $\langle s \rangle_{-+}$ across the same range of quench times, we obtain the scalings
\[
\langle s \rangle_{+-} \propto \tau_Q^{0.27 \pm 0.043},
\qquad
\langle s \rangle_{-+} \propto \tau_Q^{0.27 \pm 0.043}.
\]

These results demonstrate that the conditioned spacing statistics also follow universal KZ dynamics, consistent with the theoretical exponent $\nu/(1+z\nu)=1/4$ for mean-field critical exponents $z=2$ and $\nu=1/2$, as shown in Fig.~\ref{spacing_winding_kzm}(e).
% Also, the mean spacing corresponds to the distribution $P_{+-}(S) = P_{-+}(S)$ satisfy the relation $\langle s \rangle_{+-} = \langle s \rangle_{-+} = \sqrt{2} \langle s \rangle$. Similarly, we calculate the $\langle s \rangle_{+-}$ and $\langle s \rangle_{-+}$ for different quench times $\tau_Q$ for quantized vortices of both the phases independently and find the universal KZ scaling as: $\langle s \rangle_{+-} \propto \tau_Q^{0.29 \pm 0.029}$ and $\langle s \rangle_{-+} \propto \tau_Q^{0.29 \pm 0.029}$. These fittings reveal the universal dynamics characterizes by the mean spacing and are consistent with the theoretically estimated power law exponent $\nu/1+z\nu = 1/4$ shown in the Fig.~\ref{spacing_winding_kzm}(e) with the mean field exponents $z=2$ and $\nu=1/2$. Error bars set the width of the standard deviation across many realizations.
Furthermore we evaluate the conditioned mean spacing of FQVs based on their effective topological charge arising from vortex pairing. For different quench times $\tau_Q$, replacing $\langle s \rangle$ in Eq.~\eqref{conditioned_spacing_kz} by $\langle s_f \rangle$, the conditioned spacings follow the power-law scalings
$\langle s_f \rangle_{++} \propto \tau_Q^{0.27 \pm 0.037}$,
$\langle s_f \rangle_{--} \propto \tau_Q^{0.28 \pm 0.039}$,
$\langle s_f \rangle_{+-} \propto \tau_Q^{0.29 \pm 0.052}$, and
$\langle s_f \rangle_{-+} \propto \tau_Q^{0.28 \pm 0.053}$,
as shown in Fig.~\ref{spacing_winding_kzm}(f). 
% Besides the independent calculation of vortices from both the phases, we characterize the mean spacing of FQVs with positive winding and negative winding depending on the vortex pairing. We similarly calculate the mean spacing for different quench times $\tau_Q$ and estimate the power law scaling as: $\langle s_f \rangle_{++} \propto \tau_Q^{0.28 \pm 0.035}$, $\langle s_f \rangle_{--} \propto \tau_Q^{0.28 \pm 0.036}$, $\langle s_f \rangle_{+-} \propto \tau_Q^{0.29 \pm 0.050}$, and $\langle s_f \rangle_{-+} \propto \tau_Q^{0.29 \pm 0.052}$ shown in the Fig.~\ref{spacing_winding_kzm}(d). 
% The deviation from the theoretical model for fast quench regime indicates the breakdown of KZM. We have shown that the mean spacing of vortices independently calculated from both the phases and for the FQVs deviates in the fast quench regime as vortices from both phases does not pair up to form a FQV due to the large quantum fluctuations and some vortices remain as HQVs. The effective fitting range for $\langle s \rangle$ changes for FQVs from the fitting of mean spacing for quantized vortices as we characterize them independently in both the phases. 

\subsection{Cell size statistics via Poisson Voronoi tesselation}
Voronoi tessellation is a widely used tool in stochastic geometry for characterizing spatial point patterns~\cite{Ferenc_2007, chiu_2013}. It has also been applied to study the spatial organization of topological defects~\cite{Reichhardt_2003}. For uncorrelated points, the resulting structures are known as Poisson--Voronoi (PV) diagrams~\cite{gonzalez_2011}, which arise in many physical systems~\cite{weaire_1984, Mulheran_2000}. Since the spontaneously generated vortices in our system follow a PPP, PV tessellation provides a natural framework to analyze their spatial structure.

In two dimensions, Voronoi tessellations have been widely used to study vortex statistics and KZ dynamics~\cite{Reichhardt_2003, jimenez_2021, Thudiyangal_2024, Reichhardt_2022, Stoop_2018}. We construct Voronoi diagrams using vortex positions as centers for both spin components. 
%A representative example for the $j=1$ component is shown in Fig.~\ref{voronoi_panel}(a).
Given a set of centers $\{\mathbf{x}_i\}$ distributed over a two-dimensional domain $\mathcal{D}$, the Voronoi cell $V_i$ associated with center $\mathbf{x}_i$ contains all points $\mathbf{x}$ that are closer to $\mathbf{x}_i$ than to any other center. This condition can be written as
\begin{equation}
V_i = \{ \mathbf{x} \in \mathcal{D} \;|\; \ell(\mathbf{x},\mathbf{x}_i) \le \ell(\mathbf{x},\mathbf{x}_j), \; \forall j\neq i \},
\end{equation}
where $\ell(\mathbf{x},\mathbf{x}_i)=||\mathbf{x}-\mathbf{x}_i||$ denotes the Euclidean distance.
An important quantity characterizing the tessellation is the distribution of Voronoi cell areas. If $A_i$ denotes the area of the $i^{\rm th}$ cell, we study the scaled area $A=A_i/\langle A\rangle$, where $\langle A\rangle$ is the mean cell area. In our simulations, Voronoi cells are constructed for vortex configurations from both spin components independently, while cells extending beyond the simulation domain $[-L_x,L_x]\times[-L_y,L_y]$ with $L_x=L_y=15$ are excluded. As a representative example, Figure~\ref{voronoi_panel}(a) illustrates the tessellation with colored polygons for the $j=1$ component.
The resulting area distributions $P(A_1)$ and $P(A_2)$, obtained from many stochastic realizations, are shown in Fig.~\ref{voronoi_panel}(b) and Fig.~\ref{voronoi_panel}(c) for the two components. It is worth noting that for dimensions $D>1$, there is no exact analytical expression for the PV cell-area distribution; most results rely on numerical studies and empirical fits~\cite{gonzalez_2011}. However, the distribution is well approximated by a Gamma distribution~\cite{weaire_1986, DiCenzo_1989, Ferenc_2007}:
\begin{equation}
P(A)=\frac{b^a}{\Gamma(a)}A^{a-1}\exp(-bA).
\label{gamma}
\end{equation}

\begin{figure}[!htbp]
    \centering
    \includegraphics[width=0.9\linewidth]{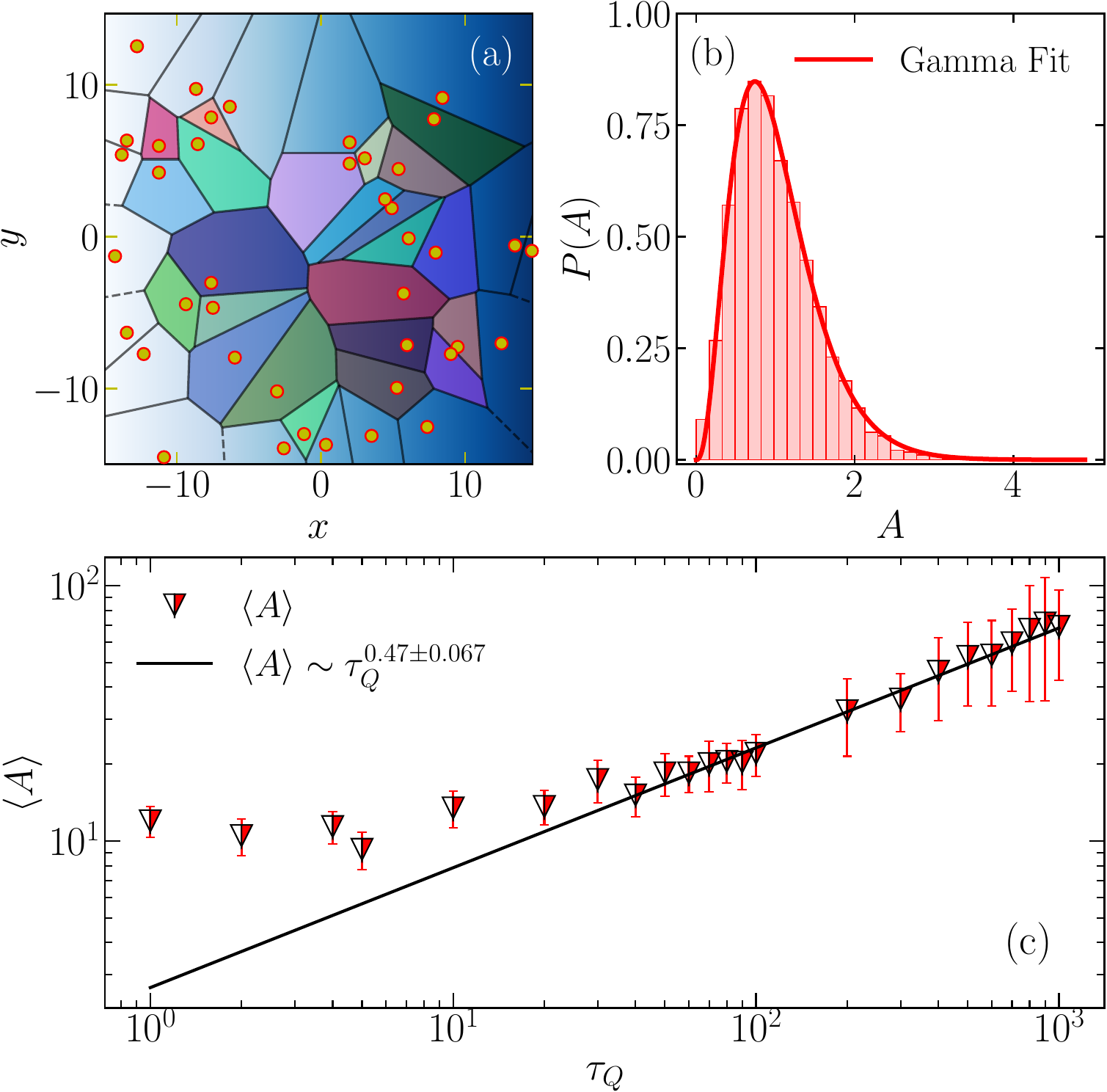}
    \caption{Voronoi tessellation of full quantum vortices (FQVs) at $t \sim t_{\rm eq}$ for $\tau_Q=50$. 
(a) Poisson--Voronoi diagram constructed using the FQV coordinates (yellow dots). 
(b) Voronoi cell-area distribution $P(A)$ obtained from a 30-bin histogram over $\mathcal{R}=500$ stochastic realizations; the solid curve shows the gamma distribution of Eq.~\eqref{gamma}. 
(c) Mean Voronoi cell area versus quench time $\tau_Q$. Weighted least-squares fits over $\tau_Q=40$--$1000$ yield the Kibble--Zurek scaling $\langle A\rangle \propto \tau_Q^{0.47\pm0.067}$, consistent with Eq.~\eqref{voronoi_pl}. Error bars denote one standard deviation over $\mathcal{R}=100$ realizations.}
    \label{voronoi_fqv}
\end{figure}
In our simulations, the best-fit parameters are $a=3.68$, $b=3.65$ for the $j=1$ component and $a=3.67$, $b=3.65$ for $j=2$. These values are in excellent agreement with previous studies of PV tessellations in two dimensions, where DiCenzo and Wertheim~\cite{DiCenzo_1989} reported $a=3.61$, $b=3.57$, and Weaire \textit{et al.}~\cite{weaire_1986} obtained $a=b=3.61$. 

The universal scaling of the mean Voronoi cell area with the quench time $\tau_Q$ is expected to follow~\cite{Thudiyangal_2024}
\begin{equation}
    \langle A \rangle \sim \hat{\xi}^2 \propto \tau_Q^{\frac{2\nu}{1+z\nu}},
    \label{voronoi_pl}
\end{equation}
where $\hat{\xi}$ is the KZ correlation length. We compute the mean cell area for quenches in the range $\tau_Q=1$--$1000$ and obtain the scaling exponents $2\nu/(1+z\nu)=0.48\pm0.065$ for the $j=1$ component and $0.48\pm0.065$ for $j=2$, as shown in Fig.~\ref{voronoi_panel}(d). These values are consistent with the KZ prediction.
\begin{figure}[!htbp]
    \centering
    \includegraphics[width=0.9\linewidth]{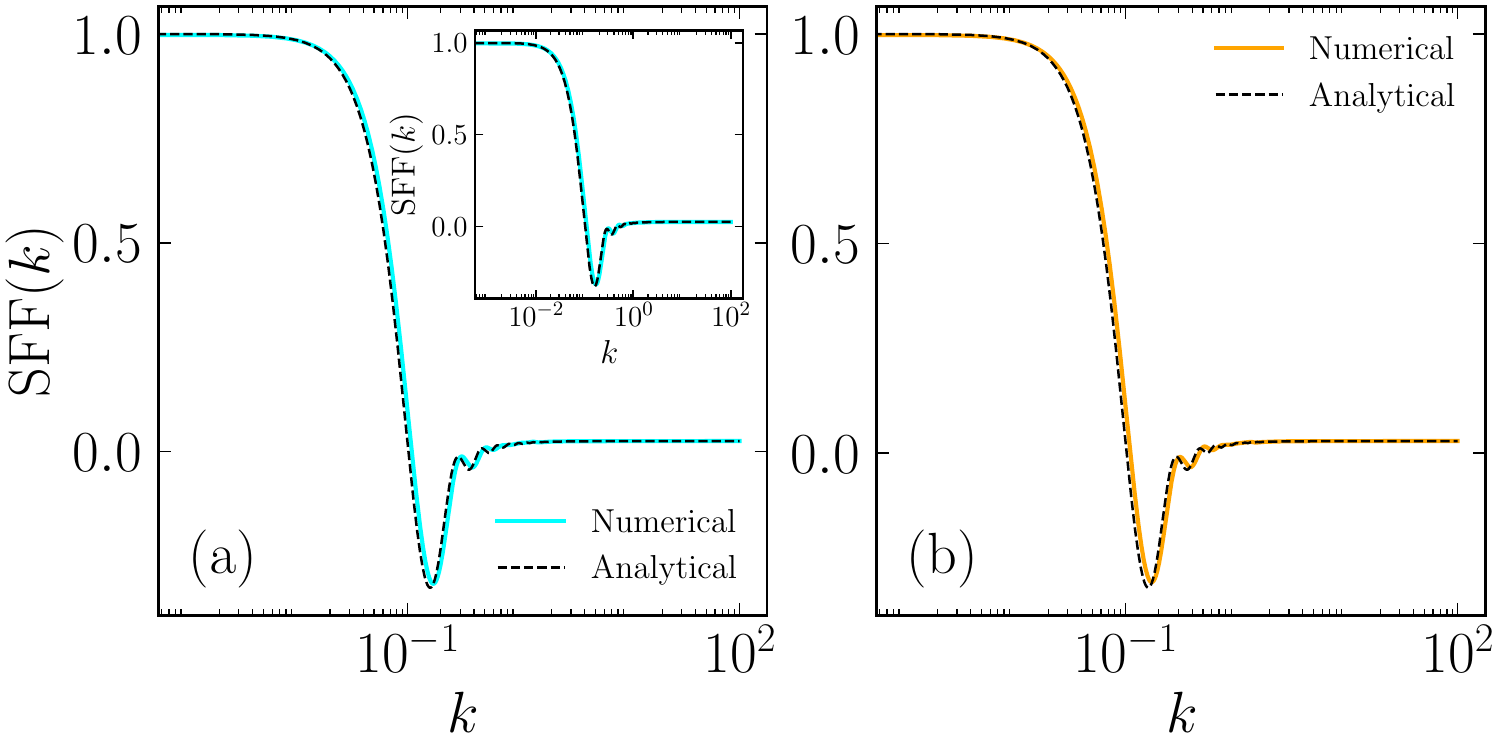}
    \caption{Spatial form factor of a newborn spinor condensate at $t \sim t_{\rm eq}$ for quench time $\tau_Q=50$. (a) SFF of vortices in the $j=1$ component without resolving topological charge; the numerical results are averaged over $\mathcal{R}=200$ realizations. The black dashed curve denotes the analytical PPP prediction [Eq.~\eqref{sff_hyper}] for $D=2$. The inset shows the corresponding SFF for the $j=2$ component. (b) SFF of full quantum vortices (FQVs), averaged over $\mathcal{R}=200$ realizations. The dashed curve corresponds to the analytical PPP result [Eq.~\eqref{sff_hyper}] for $D=2$ and $R=17.5$.}
    \label{sff_uncd}
\end{figure}

We further test the topological constraint given by Euler's formula,
$F - E + V = \chi$,
where $F$, $E$, and $V$ denote the number of faces, edges, and vertices, respectively. The Euler--Poincar\'e characteristic $\chi$ is a topological invariant that takes the value $\chi=1$ for a planar tessellation~\cite{weaire_1984}. In the thermodynamic limit, this implies an average number of edges $\langle E \rangle=6$ per cell. In our simulations, we find $\chi=1$ and obtain $\langle E \rangle=5.528(85)$ for $j=1$ and $\langle E \rangle=5.522(84)$ for $j=2$ at $t=t_{\rm eq}$ for $\tau_Q=50$. The slight deviation from the theoretical value arises from finite system-size effects.
Finally, we evaluate the second moment of the scaled cell area $y=A/\langle A\rangle$, obtaining $\langle y^2\rangle=1.264(104)$ for $j=1$ and $\langle y^2\rangle=1.264(99)$ for $j=2$, where the uncertainties are estimated from $\mathcal{R}=400$ independent realizations. These values are in good agreement with the analytical prediction $\langle y^2\rangle=1.280$ for two-dimensional Poisson--Voronoi tessellations~\cite{Gilbert_1962, Ferenc_2007}.

In addition to analyzing vortices from the two spin components independently, we also construct the PV diagram for FQVs, using their coordinates as Voronoi centers, as illustrated in Fig.~\ref{voronoi_fqv}(a) (yellow dots). The resulting cell-area distribution is well fitted by the gamma distribution of Eq.~\eqref{gamma}, with parameters $a=3.64$ and $b=3.57$, shown in Fig.~\ref{voronoi_fqv}(b). 
We further evaluate the topological properties of the tessellation and obtain the Euler characteristic $\chi=1$ with an average number of edges $\langle E\rangle=5.49(9)$. The second moment of the scaled cell area is $\langle y^2\rangle=1.264(109)$, where $y=A/\langle A\rangle$. The quoted uncertainties are estimated from $\mathcal{R}=400$ independent realizations. 
Finally, the mean cell area of the FQV PV diagram exhibits the KZ scaling $\langle A\rangle \propto \tau_Q^{0.47 \pm 0.067}$, as shown in Fig.~\ref{voronoi_fqv}(c). 

\section{Spatial form factor in a newborn spinor condensate}
\label{sff_label}
Poisson point processes (PPPs) are widely used to characterize quantum chaos~\cite{haake2010}. In random matrix theory (RMT), spectral properties of quantum chaotic systems are typically analyzed using eigenvalue spacing statistics and the spectral form factor (SFF)~\cite{Leviandier_1986, Wilkie_1991, Alhassid_1993}. Notably, the Wigner surmise also describes nearest-neighbor spacing distributions of homogeneous PPPs in $\mathbb{R}^2$~\cite{chiu_2013, Sakhr_2006}, highlighting a connection between stochastic geometry and quantum chaos. Motivated by this analogy, we employ the spatial form factor (SFF) as a probe of spatial point processes, analogous to its spectral counterpart~\cite{Wilkie_1991, Ma_1995, Campo_2017}.

Consider a set of points $E=\{r_i\}$ distributed in a domain $\mathcal{D}$, with $i=1,\dots,N$ and $r_i\in\mathcal{D}$. The SFF is defined as the even Fourier transform of pairwise distances~\cite{Massaro25}:
\begin{equation}
{\rm SFF}(k)=\left\langle\frac{1}{N^2}\sum_{i,j=1}^{N}\cos\big(k|r_i-r_j|\big)\right\rangle,
\label{sff}
\end{equation}
where $k$ is the wave vector and $\langle\cdot\rangle$ denotes averaging over configurations. This form closely parallels the spectral form factor defined from eigenvalue spacings~\cite{mehta2004}.

Separating diagonal ($i=j$) and off-diagonal ($i\neq j$) contributions and using the earlier definition of the pair distance between defects as $s=|r_i-r_j|$, Eq.~(\ref{sff}) can be rewritten as
\begin{equation}
{\rm SFF}(k)=\frac{1}{N}+\left(1-\frac{1}{N}\right)\int_{0}^{s_{\rm max}}\cos(ks)\,P_{\mathcal{D}}(s)\,ds,
\label{sff_numeric}
\end{equation}
where $P_{\mathcal{D}}(s)$ is the pair-distance distribution (see Sec.~\ref{subsubsec:poissonPP}). 

For a binomial point process (BPP) in a $D$-dimensional ball of radius $R$, $P_{\mathcal{D}}(s)$ is given by the ball line-picking distribution~\cite{kendall1963, santalo2004}, yielding the analytical result~\cite{Massaro25}
\begin{widetext}
\begin{equation}
{\rm SFF}(k)=\frac{1}{N}+\left(1-\frac{1}{N}\right)
\,{}_2F_3\!\left(\frac{D+1}{2},\frac{D}{2};\frac{1}{2},\frac{D+2}{2},D+1;-k^2R^2\right),
\label{sff_hyper}
\end{equation}
\end{widetext}
where ${}_2F_3$ is the generalized hypergeometric function~\cite{Gradshteyn1996}. For PPPs, the SFF follows by averaging the BPP expression over realizations with varying $N$.

In our system, vortices are generated at random positions and are well described by a PPP (see Sec.~\ref{sub:unconditioned_dist}). We therefore compute the SFF numerically using Eq.~(\ref{sff_numeric}), based on the measured $P_{\mathcal{D}}(s)$, and average over $\mathcal{R}=200$ realizations. Figure~\ref{sff_uncd}(a) shows the resulting SFF for vortices in the $j=1$ component without resolving their topological charge. The SFF exhibits the characteristic dip–ramp–plateau structure familiar from quantum chaotic systems.
We compare the numerical results with the analytical PPP prediction in Eq.~(\ref{sff_hyper}) for $D=2$ and $R=17.5$, shown as a dashed curve in Fig.~\ref{sff_uncd}(a). In the analytical expression, $1/N$ is replaced by its ensemble average $\langle 1/N\rangle$, where $N$ denotes the number of vortices within the disk of radius $R$. The numerical and analytical results show excellent agreement.
In addition to vortices from individual components, we also evaluate the SFF for FQVs formed by pairing vortices from complementary phases. Ignoring phase circulation, we compute the numerical SFF averaged over $\mathcal{R}=200$ realizations, as shown in Fig.~\ref{sff_uncd}(b). The results are in excellent agreement with the analytical PPP prediction in Eq.~\eqref{sff_hyper} for $D=2$ and $R=17.5$.

\begin{figure*}[!htbp]
    \centering
    \includegraphics[width=0.8\linewidth]{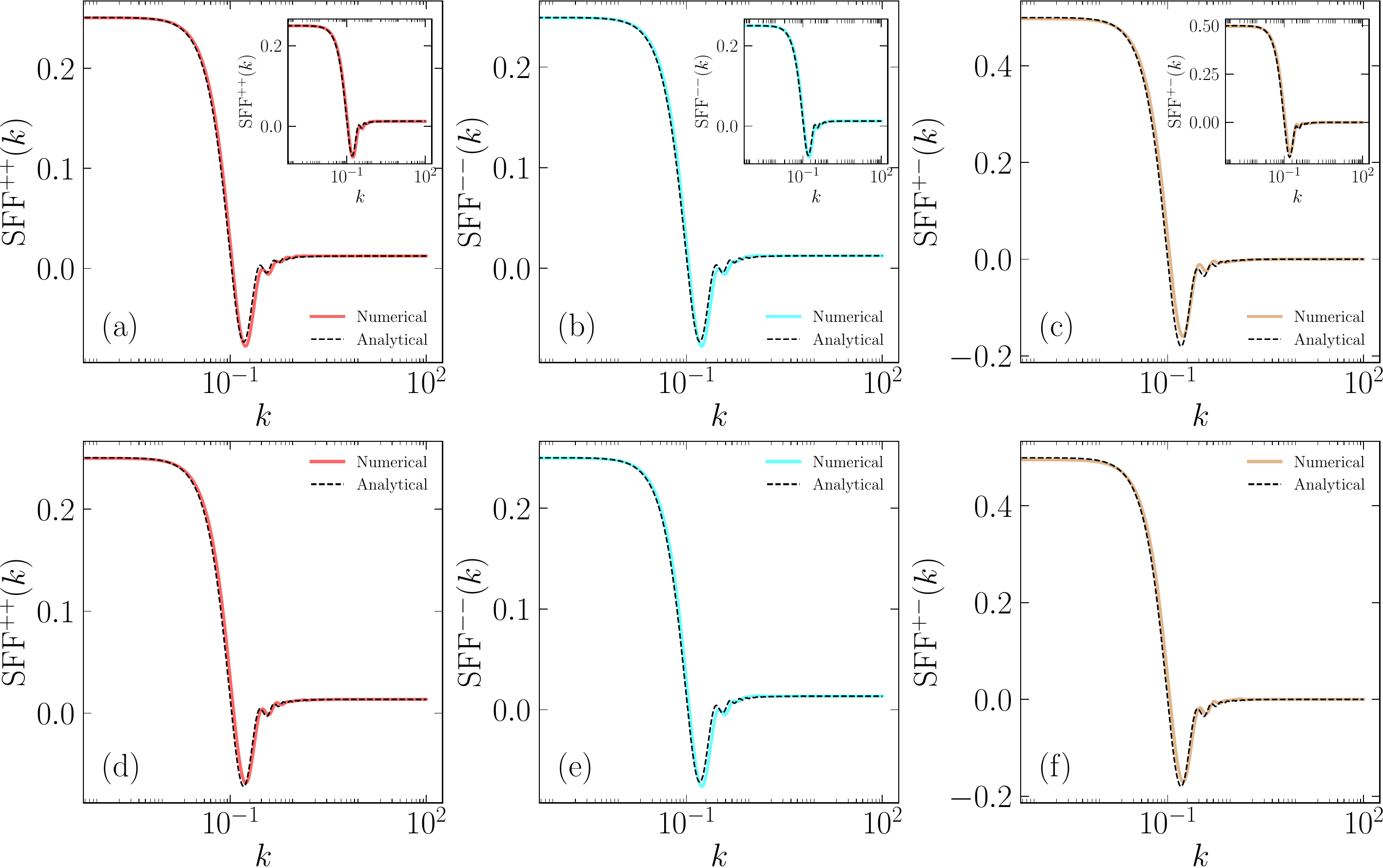}
    \caption{Spatial form factor of vortices resolved by topological charge at $t \sim t_{\rm eq}$ for quench time $\tau_Q=50$. Panels (a)–(c) show the SFF of quantized vortices in the $j=1$ component (insets: $j=2$). Panel (a) corresponds to vortices with positive winding, (b) to negative winding, and (c) to opposite-charge correlations. The black dashed curves denote the analytical PPP predictions [Eqns.~\ref{analytic_sff_winding1}–\ref{analytic_sff_winding3}] for $D=2$ and $R=17.5$. All numerical results are averaged over $\mathcal{R}=200$ realizations. 
Panels (d)–(f) show the corresponding SFF for FQVs resolved by charge. The numerical results, averaged over $\mathcal{R}=200$ realizations, are in excellent agreement with the analytical PPP expressions [Eqns.~\ref{analytic_sff_winding1}–\ref{analytic_sff_winding3}].}
    \label{sff_charge}
\end{figure*}
We now extend our analysis by incorporating the topological winding of vortices into the SFF. The SFF can be decomposed into contributions from vortices with definite charge labels as
\begin{equation}
{\rm SFF}(k) = {\rm SFF}^{++}(k) + {\rm SFF}^{--}(k) + {\rm SFF}^{+-}(k).
\end{equation}

Defining
\begin{equation}
F^{cc'}(k)=\frac{1}{N^2}\left\langle \sum_{i=1}^{N_c}\sum_{j=1}^{N_{c'}} \cos\big(k|r_i-r_j|\big) \right\rangle,
\end{equation}
with $c,c'\in\{+,-\}$ denoting vortex charges, the contributions are given by ${\rm SFF}^{++}(k)=F^{++}(k)$, ${\rm SFF}^{--}(k)=F^{--}(k)$, and ${\rm SFF}^{+-}(k)=2F^{+-}(k)$. Here, ${\rm SFF}^{++}$ and ${\rm SFF}^{--}$ describe correlations between vortices of the same charge, while ${\rm SFF}^{+-}$ captures correlations between oppositely charged vortices.

For PPPs, the corresponding analytical expressions are
\begin{equation}
{\rm SFF}^{++}(k)=\left\langle\frac{N_+}{N^2}\right\rangle + \left\langle \frac{N_+(N_+-1)}{N^2} \right\rangle \mathcal{I}_D,
\label{analytic_sff_winding1}
\end{equation}
\begin{equation}
{\rm SFF}^{--}(k)=\left\langle\frac{N_-}{N^2}\right\rangle + \left\langle \frac{N_-(N_--1)}{N^2} \right\rangle \mathcal{I}_D,
\label{analytic_sff_winding2}
\end{equation}
and
\begin{equation}
{\rm SFF}^{+-}(k)=2\left\langle \frac{N_+N_-}{N^2} \right\rangle \mathcal{I}_D,
\label{analytic_sff_winding3}
\end{equation}
where $\mathcal{I}_D = {}_2F_3\!\left(\frac{D+1}{2},\frac{D}{2};\frac{1}{2},\frac{D+2}{2},D+1;-k^2R^2\right)$, and $\langle\cdot\rangle$ denotes ensemble averaging.

We first analyze quantized vortices in each spin component independently and compute the SFF numerically using the pair-distance distribution $P_{\mathcal{D}}(s)$, resolved by vortex charge. The results, averaged over $\mathcal{R}=200$ realizations, are shown in Fig.~\ref{sff_charge}(a)-(c) for ${\rm SFF}^{++}(k)$, ${\rm SFF}^{--}(k)$, and ${\rm SFF}^{+-}(k)$ in the $j=1$ component, with the corresponding results for $j=2$ shown in the insets. The numerical results are in excellent agreement with the analytical PPP predictions.
We further compute the SFF for FQVs, classified by their charge labels, as shown in Fig.~\ref{sff_charge}(d)-(f). The numerical results again show excellent agreement with the analytical expressions for PPPs.

We further compute the SFF for FQVs, classified by their charge labels, as shown in Fig.~\ref{sff_charge}(d)-(f). The numerical results again show excellent agreement with the analytical expressions for PPPs.

\section{\label{sec:conclusions}Conclusions}

In this work, we have investigated the nonequilibrium condensation of a homogeneous 
coherently coupled Bose-Bose mixture across a continuous symmetry-breaking 
transition. Consistent with the Kibble-Zurek mechanism, we find that the 
equilibration time and the density of defects obey universal scaling laws governed 
by the quench rate. Our findings also illustrate that the spatial organization of 
the defects exhibit robust universal features beyond the standard Kibble–Zurek 
scaling of defect density. Despite the presence of both coherent and 
interaction-mediated coupling between the two condensate components, the spatial 
distribution of these vortices is well described by a two-dimensional homogeneous Poisson 
point process around the equilibration time. In contrast to a single-component 
condensate, vortices in one component are intrinsically coupled to the dynamics 
of vortices in the other component through both the interspecies interaction 
and the Rabi coupling. One might therefore expect significant intercomponent 
correlations capable of invalidating the assumption of independently distributed 
defects. Instead, we find that the emergent vortex configurations retain the universal 
spatial statistics reminiscent of a Poisson point process. This universality is 
supported by several indicators, including pair-distance statistics, Voronoi 
tessellation, and the spatial form factor, which displays a characteristic 
dip–ramp–plateau structure analogous to spectral correlations in quantum chaotic 
systems. Importantly, the same statistical framework remains applicable when the 
elementary defects bind as full quantum vortices, demonstrating that 
the emergence of composite topological objects does not alter the 
underlying spatial universality.

Our predictions are directly accessible in experiments with ultracold atomic mixtures 
confined in homogeneous traps, which provide an ideal platform for probing universal 
nonequilibrium dynamics without the causality-induced corrections that often arise 
in inhomogeneous systems~\cite{Zurek_2009,delCampo2013,Pyka13,Ulm13,KimShin22}. The 
universal spatial statistics of topological defects identified here may also have 
implications beyond the Kibble--Zurek scenario, particularly for the spontaneous 
emergence of quantum turbulence following a 
quench~\cite{Shinn2025SQT,Massaro25migdal} driving the phase transition. 
In turbulent multicomponent superfluids, the incompressible kinetic-energy spectrum 
is intimately linked to the spatial arrangement of vortices in each component, 
motivating the extension of existing theoretical 
frameworks~\cite{Novikov1975,Chavanis2001,Bradley2012QT} to coherently coupled 
mixtures. Another promising direction is the investigation of 
Migdal's area law in quantum turbulence~\cite{MIGDAL2023,Massaro25migdal} and spin
domains~\cite{Saito_2007}. Such studies may help establish spatial defect statistics as a unifying 
framework for understanding the geometry and dynamics of vortices across a broad 
range of nonequilibrium quantum fluids.

\begin{acknowledgments}
We thank Sivasankar  and Sunilkumar Venkateshappa for
several insightful discussions. 
 A.Roy acknowledges the support of the Science and Engineering Research Board (SERB),
Department of Science and Technology, Government of India, under the project
SRG/2022/000057 and IIT Mandi seed-grant funds under the project IITM/SG/AR/87.
A.Roy acknowledges the National Supercomputing Mission (NSM) for providing
computing resources of PARAM Himalaya at IIT Mandi, which is implemented by
C-DAC and supported by the Ministry of Electronics and Information Technology
(MeitY) and Department of Science and Technology (DST), Government of India.
P.C acknowledges support from “Quantum Optical Networks based on Exciton-polaritons” (Q-ONE, N. 101115575, HORIZON-EIC-2022-PATHFINDER CHALLENGES EU project), "Neuromorphic Polariton Accelerator" (PolArt, N.101130304, Horizon-EIC-2023-Pathfinder Open EU project), “National Quantum Science and Technology Institute” (NQSTI, N. PE0000023, PNRR MUR project), “Integrated Infrastructure Initiative in Photonic and Quantum Sciences” (I-PHOQS, N. IR0000016, PNRR MUR project).
\end{acknowledgments}

 \let\itshape\upshape
 \normalem
 \bibliography{reference}
%-------------------------------------
\newpage
\pagebreak
\clearpage
%\onecolumngrid
% \begin{center}
% \textbf{\large Supplementary Material for "Kibble-Zurek Scaling and Spatial Statistics in Quenched Binary Bose Superfluids"}
% \end{center}
% \begin{abstract}
%     abcd
% \end{abstract}

\onecolumngrid
% \section{ABCD}

%\section*{Appendix}
\appendix
% \section*{Appendix}
% \setcounter{section}{0}
\begin{center}
\
\end{center}

\section{Characterization of the equilibration time}
\label{charc_t_eq}
\begin{figure}[!htbp]
    \centering
    \includegraphics[width=0.7\linewidth]{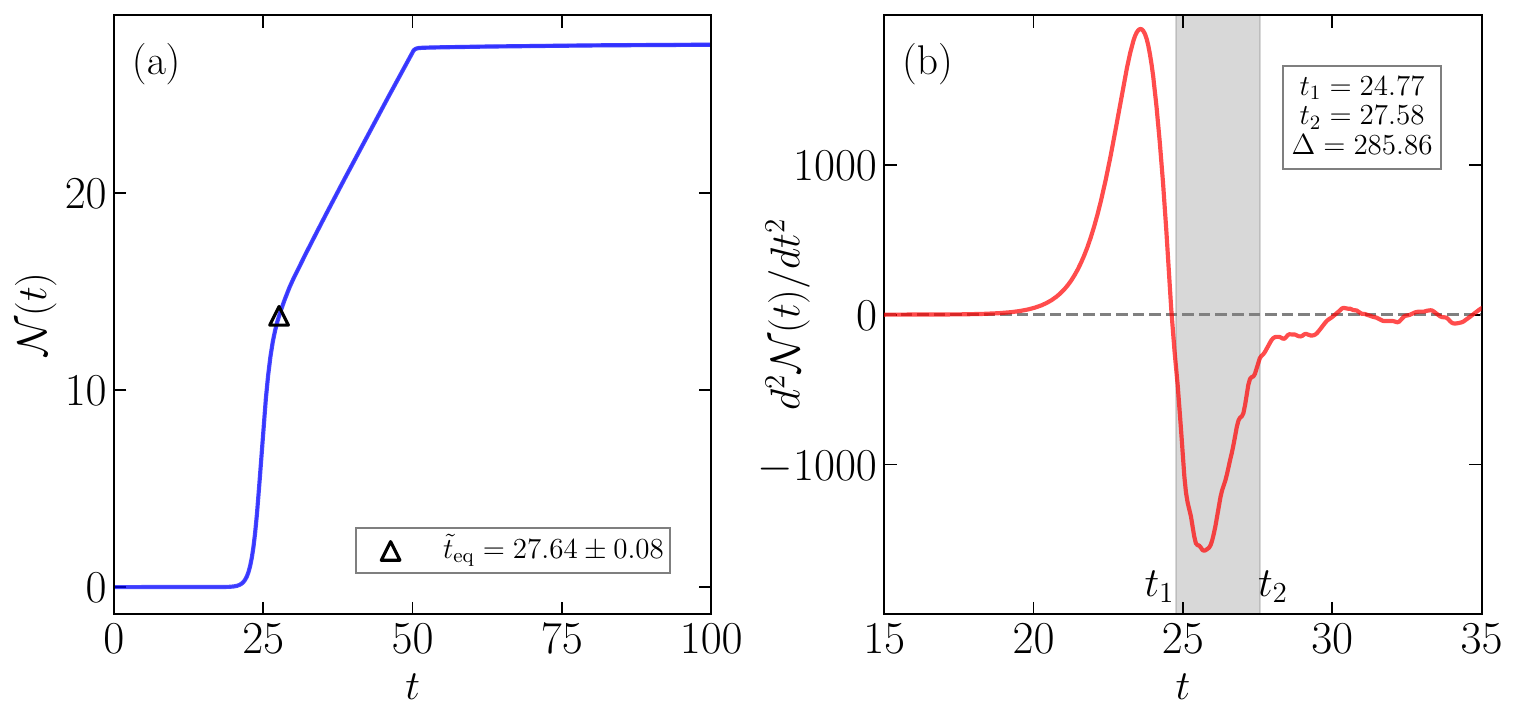}
    \caption{Determination of $t_{\rm eq}$. Panel (a) shows the temporal evolution of the averaged condensate norm $\mathcal{N}(t)$ for the quench time $\tau_Q=50$. The triangular marker represents the equilibration time $t_{\rm eq}$ averaged over $\mathcal{R}=100$ independent trajectories. Panel (b) represents the second derivative of the norm $\mathcal{N}(t)$ for a single realization. The shaded region represents the time interval $(t_1, t_2)$ for $d^2\mathcal{N}(t)/dt^2 \le - \Delta$, where $\Delta$ is set to be $15\%$ of $max(d^2\mathcal{N}(t)/dt^2)$.}
    \label{norm_t_eq}
\end{figure}

The equilibration time $t_{\rm eq}$ is determined from the evolution of the condensate norm $\mathcal{N}(t) = {(1/\mathcal{A})}\sum_{j=1,2}\int d\mathbf{x}\, |\psi_j(\mathbf{x},t)|^2$ shown in the Fig.~\ref{norm_t_eq}(a). Motivated by \cite{chesler_2015, Shinn2025SQT}, we calculate the maximum of $d^2\mathcal{N}(t)/dt^2$ and take $15 \%$ of the maximum as $\Delta$. We then identify the time interval $(t_1, t_2)$ that satisfies $d^2\mathcal{N}(t)/dt^2 \le - \Delta$ and set $\tilde t_{\rm eq} = t_2$ illustrated in the Fig.~\ref{norm_t_eq}(b). Note that in the SPGPE framework, the critical chemical potential satisfies $\mu_c \neq 0$, so the system does not cross criticality exactly at $t=0$~\cite{Liu_2020}. While this introduces a small shift in the reference time for freeze-out, we find that the resulting corrections do not affect the scaling behavior, which remains governed by the standard KZ power law. To account for this offset, we define an effective equilibration time, relative to the time $t_{\rm c}$ when the system crosses the critical point. Typically, the freeze-out time $\hat t$ is measured relative to the critical crossover $t_{\rm c}$ at $\mu_c := \mu(t_{\rm c})$, and the equilibration time is proportional to the freeze-out time~\cite{chesler_2015}
\begin{equation}
     t_{\rm eq} := \tilde t_{\rm eq} - t_{\rm c} \propto \hat t \propto \tau_Q^{z\nu/1+z\nu},
\end{equation}
where $t_{\rm c}=c\tau_Q$ with $c = (\mu_c - \mu_i)/(\mu_f - \mu_i)$. 
Our numerical $t_{\rm eq}$ scales with $\tau_Q$ as
\begin{equation}
    t_{\rm eq} = (4.33 \pm 0.135)\tau_Q^{0.47\pm0.008} + (0.0127\pm0.0032)\tau_Q.
\end{equation}
% \paolo{why c is negative? this cannot be, $t_{\rm c}$ cannot be negative}
We then use the shifted time $t_{\rm eq} := (\tilde t_{\rm eq} - c\tau_Q)$ as the natural time scale to study the KZ defect density.

\section{Identification of full quantum vortex}
\label{fqv_threshold}

\begin{figure}
    \centering
    \includegraphics[width=0.55\linewidth]{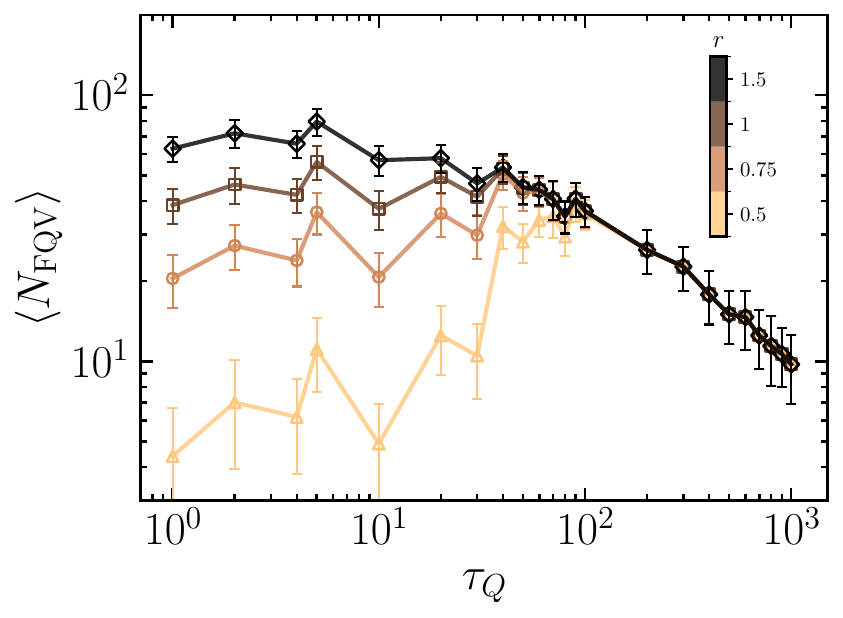}
    \caption{Counting full quantum vortices (FQVs) $N_{\rm FQV}$ for the quench times $\tau_Q = 1-1000$. FQVs are identified as composite defects formed by pairing vortices with same topological winding number from both phases within a spatial cutoff $r$. We count FQVs for different $\tau_Q$ values ranging from $1$ to $1000$ with a set of cutoff values $r = [0.5, 0.75, 1.0, 1.5]$. Each data point is averaged over $\mathcal{R} = 100$ noise trajectories. The error bars represent one standard deviation across those realizations.}
    \label{fqv_r}
\end{figure}

We identify the full quantum vortices (FQVs) as composite defects formed by pairing of phase circulations from the two condensate components having same toplogical winding number. Specifically, a vortex with $+2\pi$$(-2\pi)$ winding in one component pairs with a vortex having $+2\pi$$(-2\pi)$ winding from other component within a spatial distance $r$. In this way, the parameter $r$ serves as a spatial cutoff that determines whether two same sign defects are close enough to be regarded as a composite defect (FQV). In our work, we perform the analysis using various cutoff values, $r = [0.5, 0.75, 1.0, 1.5]$. Smaller $r$ values impose a stricter criterion such that only tightly bound vortex pairs are counted as composite defects i.e FQVs, whereas larger $r$ values allow us to consider weakly bound vortex pairs to be identified as FQVs. We perform the analysis for different cutoff values $r$ and quench times ranging from $\tau_Q = 1$ to $\tau_Q = 1000$, as shown in the Fig.~\ref{fqv_r}. We observe that the counting of FQVs becomes independent of the spatial cutoff $r$ for slow quenches, indicating well-bound composite defects. For faster quenches, different $r$ values yield different count since fast quenches render a higher mean defect density and makes it difficult to pair up.

\section{Spacing statistics beyond KZM}
\label{spacing_appendix}

\begin{figure}
    \centering
    \includegraphics[width=0.9\linewidth]{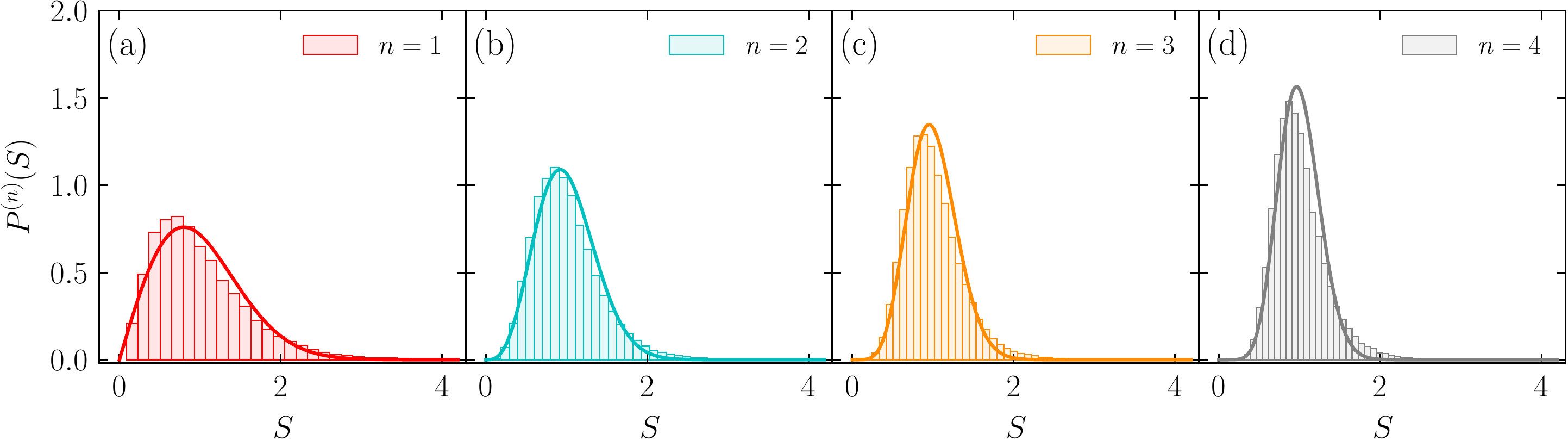}
    \caption{Nearest-neighbor (NN) spacing distributions at $t/t_{\rm eq}=1$ for the quench time $\tau_Q=50$. 
(a)--(d) NN spacing distributions $P^{(n)}(S)$ of quantized vortices in spin component $j=2$ for different NN orders. 
(a) First NN ($n=1$), where the solid curve represents the Wigner--Dyson distribution given by Eq.~\eqref{wd}. 
(b)--(d) Higher-order NN spacing distributions ($n=2,3,4$), with solid curves corresponding to the theoretical prediction for a 2D Poisson point process (PPP) given by Eq.~\eqref{nth-ND}. All histograms are constructed from $\mathcal{R}=1000$ stochastic realizations using $40$ bins.}
    \label{spacing_dist_j2}
\end{figure}
\begin{figure}
    \centering
    \includegraphics[width=0.9\linewidth]{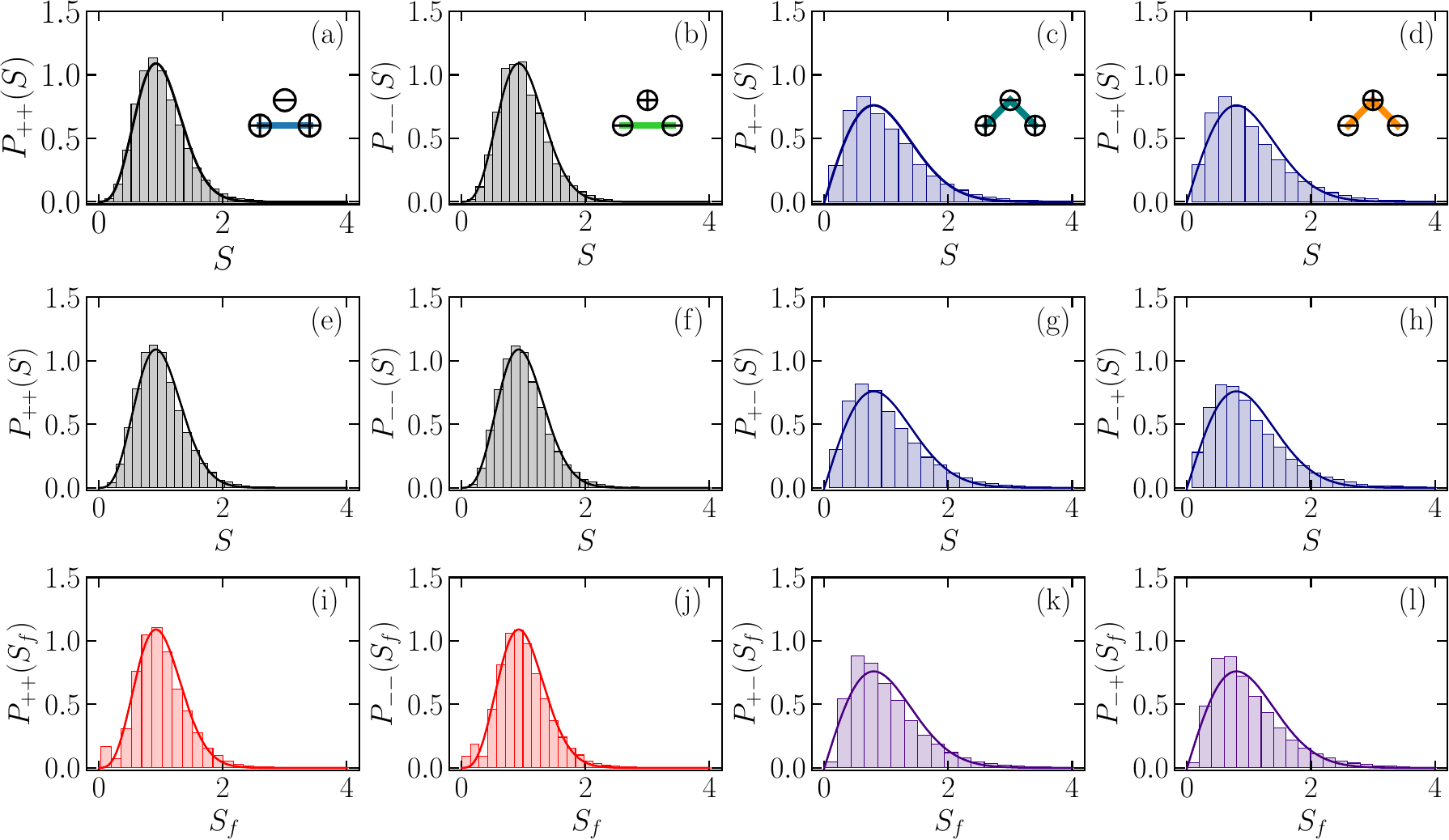}
    \caption{Universal spacing statistics conditioned on vortex topological charge at $t/t_{\rm eq}=1$ for $\tau_Q=50$. 
(a)-(d) First nearest-neighbor (NN) spacing distributions of vortices in the $j=1$ component conditioned on vortex charge: (a) $P_{++}(S)$ for positive-positive vortex pairs, (b) $P_{--}(S)$ for negative–negative vortex pairs, (c) $P_{+-}(S)$ for positive–negative vortex pairs, and (d) $P_{-+}(S)$ for negative–positive vortex pairs. Solid curves in panels (a) and (b) correspond to the theoretical prediction of Eq.~\eqref{nth-ND} with $n=2$, while those in panels (c) and (d) correspond to $n=1$ (Wigner--Dyson). Panels (e)-(h) show the corresponding NN spacing distributions of vortices in the condensate component $j=2$.
(i)-(l) First NN spacing distributions of FQVs formed by pairing two same charged vortices from complementary phases: (i) $P_{++}(S_f)$ and (j) $P_{+-}(S_f)$ with solid curves given by Eq.~\eqref{nth-ND} for $n=2$; and (k) $P_{+-}(S_f)$ and (l) $P_{-+}(S_f)$ with solid curves given by Eq.~\eqref{nth-ND} for $n=1$, respectively. All histograms are constructed from $\mathcal{R} = 1000$ stochastic realizations with 25 bins.}
    \label{spacing_charge_j_both}
\end{figure}

The NN spacing distribution of vortices without considering their topological charge is obtained within the KZM-PPP model given by the Eq.~\eqref{nth-ND} with $n$ being the NN order. The four lowest order spacing distributions that we constructed are in well agreement with the theoretical predictions shown in Figs.~\ref{nth-order_dist}(a)-(d) for vortices in condensate component $j=1$, and in Figs.~\ref{spacing_dist_j2}(a)-(d) for vortices in component $j=2$. As KZM-PPP model does not incorporate any specific form of the interaction between vortices, it can give a generic description of vortex statistics in a wide range of systems undergoing nonequilibrium symmetry-breaking transitions. Extending our study to charge resolved statistics of vortices~\cite{Thudiyangal_2024}, such as $P_{++}(S)$, constructed by considering only positively charged vortices $(w=+1)$. We calculate $1^{\rm st}$ NN distances $s$ of such vortices. Numerically, if a circle of radius $s$ is drawn around a reference vortex, the first NN vortex lies on its boundary with all other vortices lies beyond the circular region. The probability distribution is constructed using the normalized spacing $S=s/\bar s$, where $\bar s$ denotes the mean NN distance and illustrated in Fig.~\ref{spacing_charge_j_both}(a) for the vortices in condensate component $j = 1$ with the solid curve corresponds to the theoretical prediction given by the Eq.~\eqref{nth-ND} with $n=2$. The distributions $P_{--}(S)$, $P_{+-}(S)$, and $P_{-+}(S)$ are constructed in a similar fashion shown in the Fig.~\ref{spacing_charge_j_both}(b)-(d) with the solid curves correspond to Eq.~\eqref{nth-ND} with $n=2$, $n=1$, and $n=1$ respectively. The first subscript denotes the charge of the reference vortex, while the second denotes the charge of its first NN. Here, all the distributions correspond to $1^{\rm st}$ NN distances, however the distributions $P_{++}(S)$ and $P_{--}(S)$ deviate from the first NN prediction of the $n^{\rm th}$ order spacing distribution. This deviation arises because the same charged vortices repel each other, whereas vortices with opposite charges attract. As a result, a vortex is more likely to have an oppositely charged vortex as first NN than a vortex with same charge. Charge symmetry further implies $P_{++}(S)=P_{--}(S)$ and $P_{+-}(S)=P_{-+}(S)$, in agreement with our numerical estimations shown in the Fig.~\ref{spacing_charge_j_both}. Our results for FQVs follow similar NN spacing distributions, exhibiting same charge symmetry relations, and can be seen in the lower panels of the Fig.~\ref{spacing_charge_j_both}. 

\end{document}